\documentclass[a4paper]{article}
\usepackage{epsfig,amsmath,amsfonts,amssymb,setspace,multirow,textcomp}
\usepackage[T1]{fontenc}
\makeatletter
\renewcommand{\@evenfoot}{\hfil \thepage \hfil}
\renewcommand{\@oddfoot}{\hfil \thepage \hfil}
\makeatother

\renewenvironment{thebibliography}[1]{\begin{oldthebibliography}{#1}\setlength{\parskip}{0ex}\setlength{\itemsep}{0ex}}{\end{oldthebibliography}}

\begin{document}
\fontsize{11}{11}\selectfont 
\title{Unidentified aerial phenomena II. Evaluation of UAP properties}
\author{B.E.~Zhilyaev, V.\,N.~Petukhov, V.\,M.~Reshetnyk}
%
\date{\vspace*{-6ex}}
\maketitle
\begin{center} {\small $Main \,Astronomical \, Observatory, NAS \,\, of \, Ukraine, Zabalotnoho \,27, 03680, Kyiv, Ukraine$}\\
{\tt zhilyaev@mao.kiev.ua}
\end{center}

\begin{abstract}
NASA commissioned a research team to study Unidentified Aerial Phenomena (UAP), observations of events that cannot scientifically be identified as known natural phenomena. The Main Astronomical Observatory of NAS of Ukraine conducts an independent study of UAP also. For UAP observations, we used two meteor stations installed in Kyiv and in the Vinarivka village in the south of the Kyiv region. Two-side monitoring of the daytime sky led to the detection of two luminous objects at an altitude of 620 and 1130 km, moving at a speed of 256 and 78 km/s. Colorimetric analysis showed that the objects are dark: B - V = 1.35, V - R = 0.23. The size of objects is estimated to be more than 100 meters. The detection of these objects is an experimental fact. Estimates of their characteristics follow from observational data. The authors do not interpret these objects. Daytime sky monitoring in Kyiv with multi-color DSLR camera at a rate of 30 frames per second in August and September 2018 and in Vinarivka with a multi-color CMOS camera in October 2022 revealed several cases of dark objects (phantoms). The time of their existence is, as a rule, a fraction of a second. These are oval-shaped objects ranging in size from 20 to 100 meters with speeds from 2 to 30 km/s.


\end{abstract}

\section*{\sc Introduction}

The Main Astronomical Observatory of NAS of Ukraine conducts an independent study of unidentified phenomena in the atmosphere. Our astronomical work is daytime observations of meteors and space invasions. Unidentified anomalous, air, and space objects are deeply concealed phenomena. The main feature of the UAP is its extremely high speed.
The eye does not fix phenomena lasting less than one-tenth of a second. It takes four-tenths of a second to recognize an event. Ordinary photo and video recordings will also not capture the UAP. To detect UAP, we need to fine-tune (tuning) the equipment: shutter speed, frame rate, and dynamic range.

According to our data, there are two types of UAP, which we conventionally call: (1) Cosmics, and (2) Phantoms. We note that Cosmics are luminous objects, brighter than the background of the sky. Phantoms are dark objects, with a contrast, according to our data, from 50\% to several per cent. Both types of UAPs exhibit extremely high movement speeds. Their detection is a difficult experimental problem. They are a by-product of our main astronomical work, daytime observations of meteors and space intrusions.

The results of our previous UAP study are published in \cite{Zhilyaev}. Here we present some conclusions.
Flights of single and groups of objects were detected, moving at speeds from 3 to 15 degrees per second. Some bright objects exhibit regular brightness variability in the range of 10 - 20 Hz. Two-site observations of UAPs at a base of 120 km with two synchronized cameras allowed the detection of a variable object, at an altitude of 1170 km. It flashes for one-hundredth of a second at an average of 20 Hz. 

Phantom shows the color characteristics inherent in an object with zero albedos. We see an object because it shields radiation due to Rayleigh scattering. An object contrast made it possible to estimate the distance using colorimetric methods.
Phantoms are observed in the troposphere at distances up to 10 - 14 km. We estimate their size from 3 to 12 meters and speeds up to 15 km/s.

\section*{\sc Astronomical observations of bright flying objects}

\subsection*{\sc Observations}
For UAP observations, we used two meteor stations installed in Kyiv and in the Vinarivka village in the south of the Kyiv region. The distance between stations is 120 km. The stations are equipped with ASI 174 MM and ASI 294 Pro cameras, and lenses with a focal length of 50 fnd 28 mm. ASI 174 MM camera has FOV of 4.08 deg, pixel size of 24.2 arc second, actual frame rate of 46.9 fps. ASI 294 Pro camera has FOV of 9.7 deg, pixel size of 34.1 arc second, frame rate of 50 fps. 

From simple trigonometry, it is easy to determine that objects at a distance of more than 995 km will fall into the field of view of the cameras. 

The SharpCap 4.0 program was used for data recording. Observations of objects were carried out in the daytime sky. Frames were recorded in the .ser format with 8 bits. To determine the coordinates of objects, the cameras were installed in the direction of the Moon.

\subsection*{\sc Results}

Fig. 1 shows the map for detecting events using the method of blinking with a sampling frequency of 46.9 frames per second. For the record length of 5 minutes we detect two events in both cameras following each other with an interval of 0.7 sec. 

\begin{figure}[h]
\centering
\resizebox{0.45\hsize}{!}{\includegraphics[angle=000]{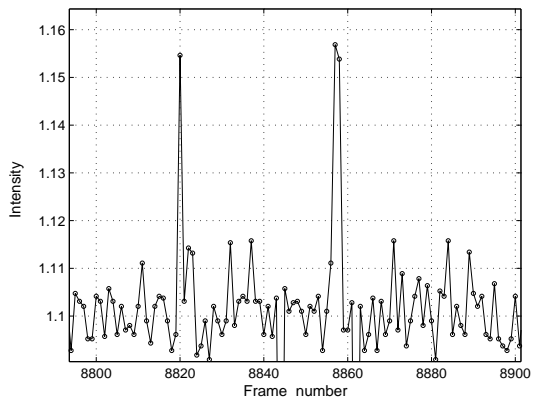}}%
\caption{The blinking map with a sampling frequency of 46.9 frames per second.} \label{figure: BrightObj_16.eps}
\end{figure}

Figs. 2 and 3 show an image with the first object taken synchronously by two cameras in 2022-10-17, 08:58:08.136 UT with time precision of one millisecond. A parallax of 0.0464 rad (2.66 degrees) gives a distance to the object of 2600 km. For an hight of 26 degrees (Stellarium, 10/17/22, 08:58:08 UT) we estimate the altitude of the object at 1130 km.
Vinarivka with 6 shots gives an angular velocity of 1.73 deg/s and a linear velocity of 78 km/s.

Figs. 4 and 5 show an image with the second object taken synchronously by two cameras in 2022-10-17,08:58:08.927 UT with time precision of one millisecond.
A parallax of 0.0848 rad (4.9 degrees) gives a distance to the object of 1400 km. For an azimuth of 280 degrees, an hight of 26 degrees, we estimate the altitude  of the object at 620 km.
Vinarivka with 10 shots gives an angular velocity of 10.27 deg/s and a linear velocity of 256 km/s.

Fig. 6, and 7 show light curves of objects. Variability is about 10 Hz.

\subsection*{\sc Evaluation of bright flying objects' properties}

\begin{figure}[!h]
\centering
\begin{minipage}[t]{.45\linewidth}
\centering
\epsfig{file = 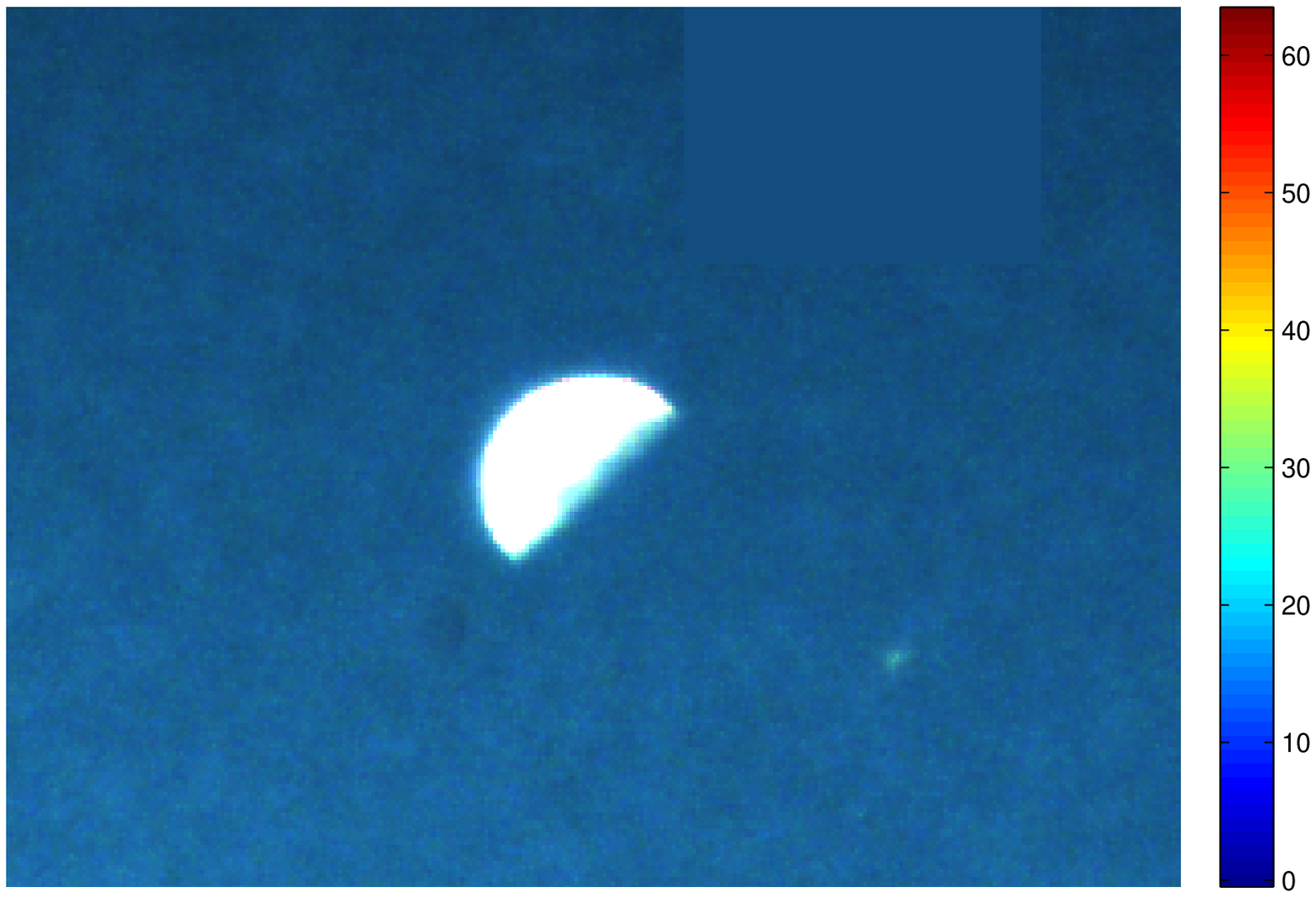,width = 1.05\linewidth} \caption{First object. Vinarivka.}\label{fig1}
\end{minipage}
\hfill
\begin{minipage}[t]{.45\linewidth} 
\centering
\epsfig{file = 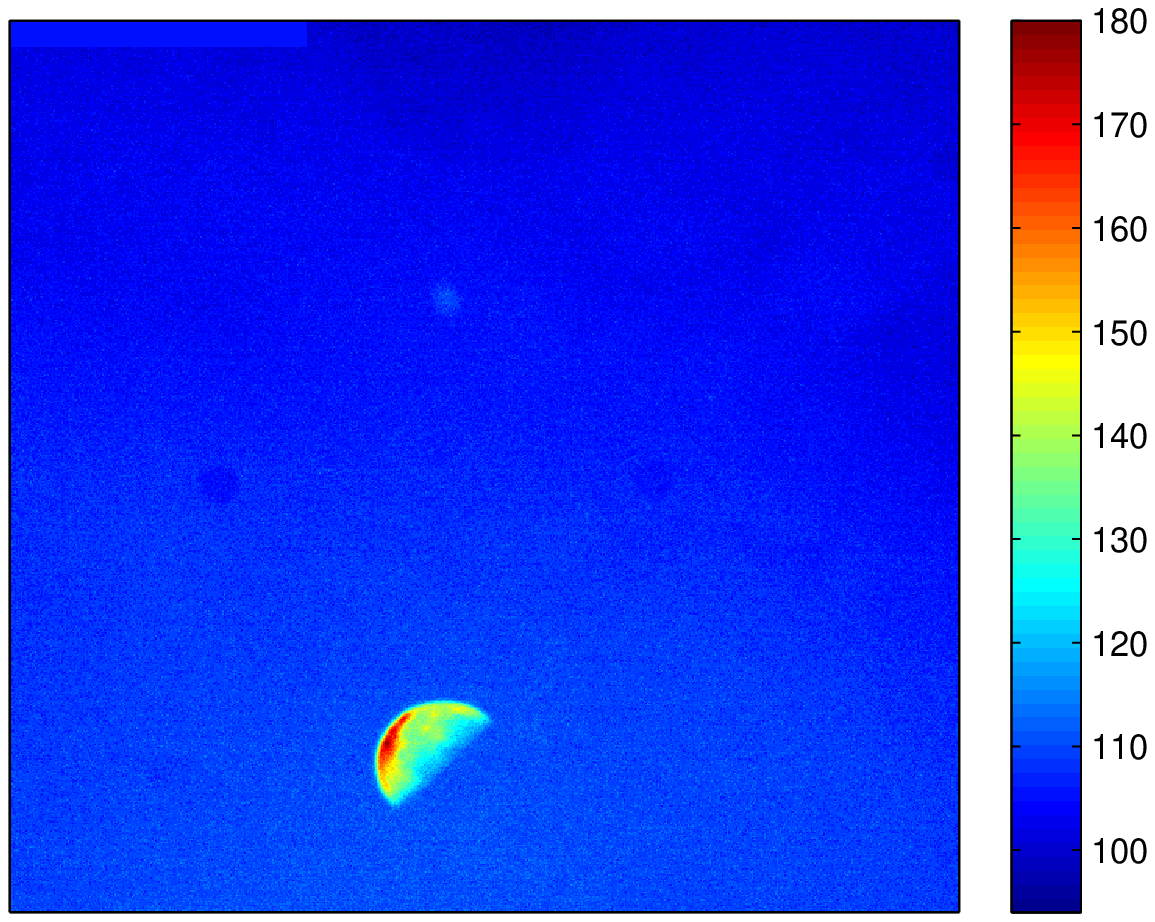,width = 0.9\linewidth} \caption{First object. Kyiv.}\label{fig2}
\end{minipage}
\end{figure}

\begin{figure}[!h]
\centering
\begin{minipage}[t]{.45\linewidth}
\centering
\epsfig{file = 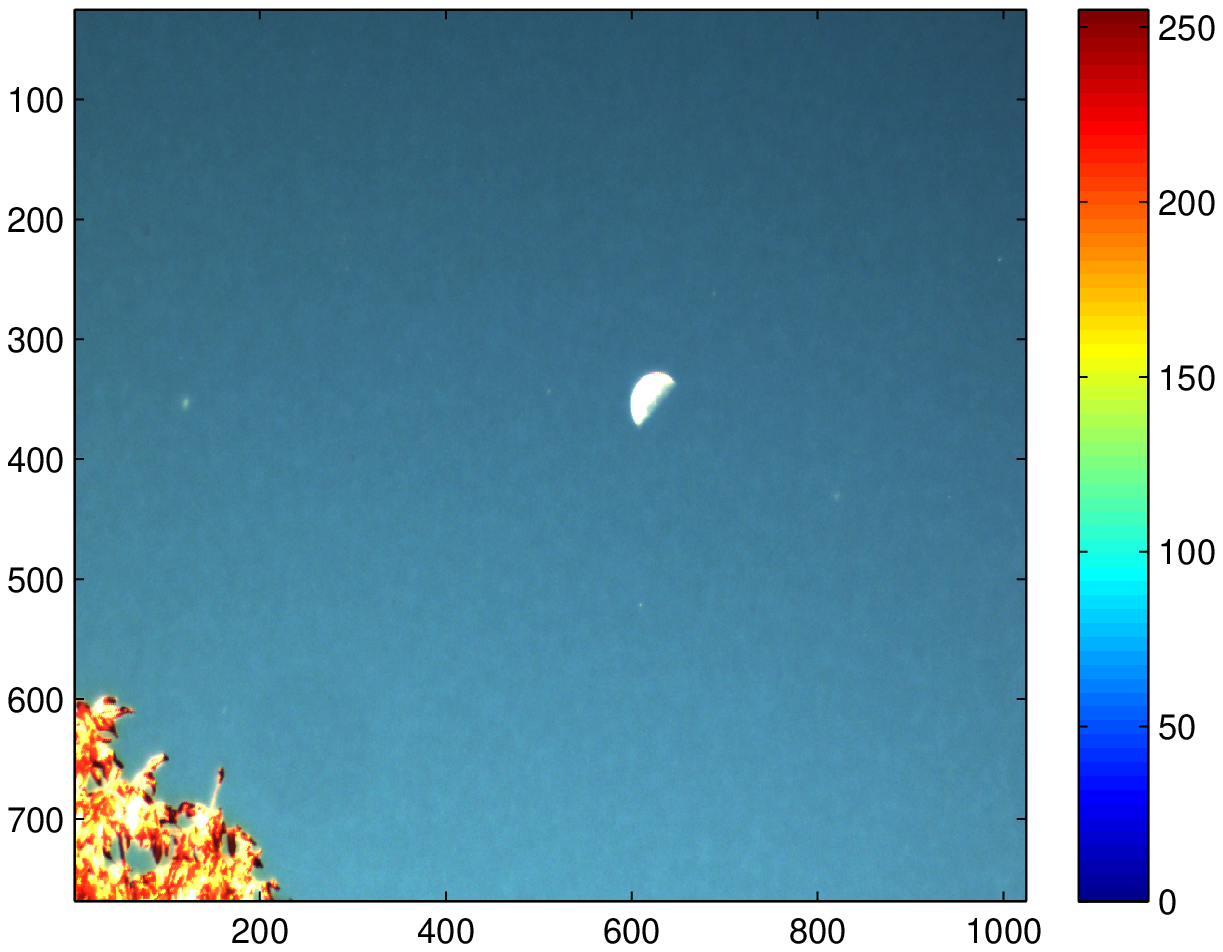,width = 1.05\linewidth} \caption{Second object. Vinarivka.}\label{fig1}
\end{minipage}
\hfill
\begin{minipage}[t]{.45\linewidth} 
\centering
\epsfig{file = 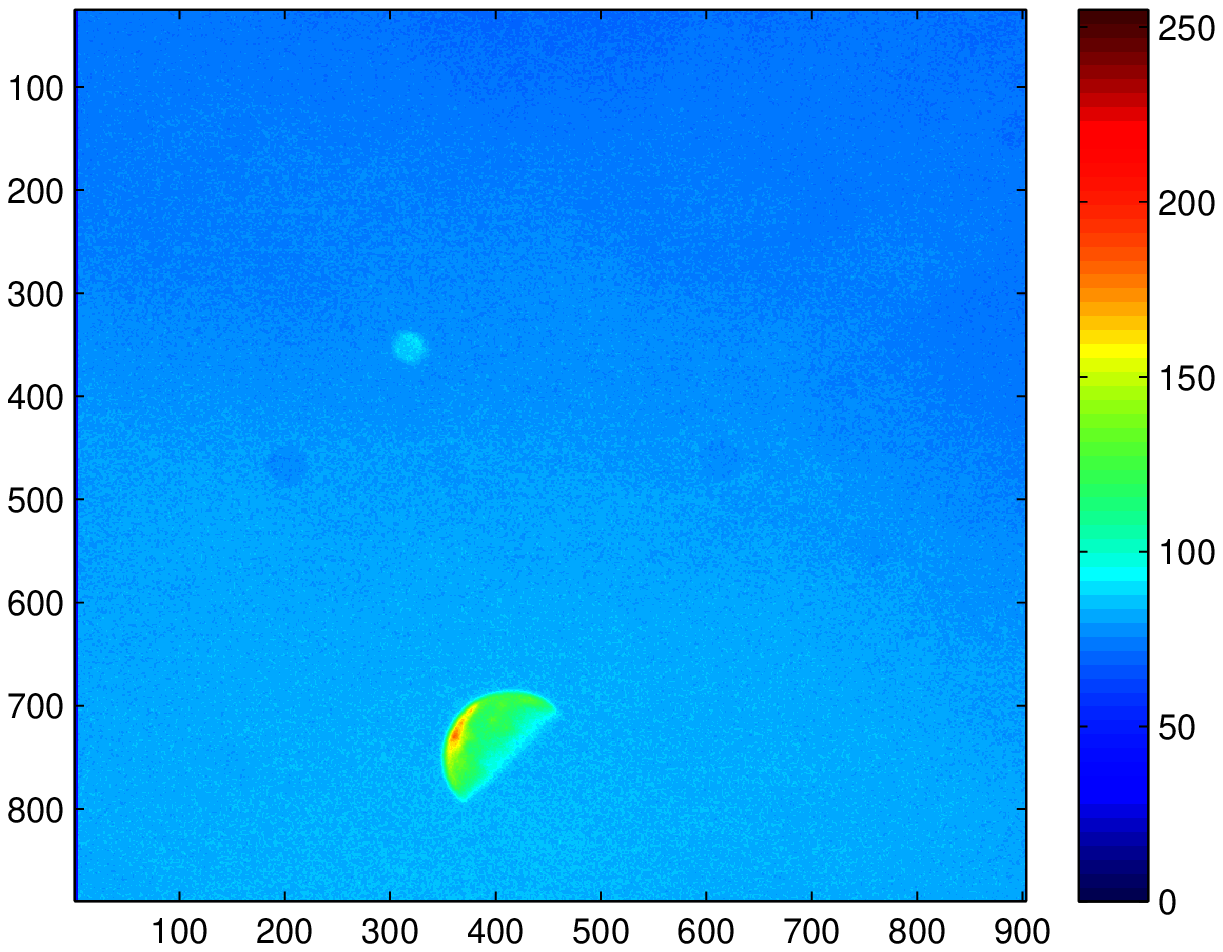,width = 1.0\linewidth} \caption{Second object. Kyiv.}\label{fig2}
\end{minipage}
\end{figure}

\begin{figure}[!h]
\centering
\begin{minipage}[t]{.45\linewidth}
\centering
\epsfig{file = 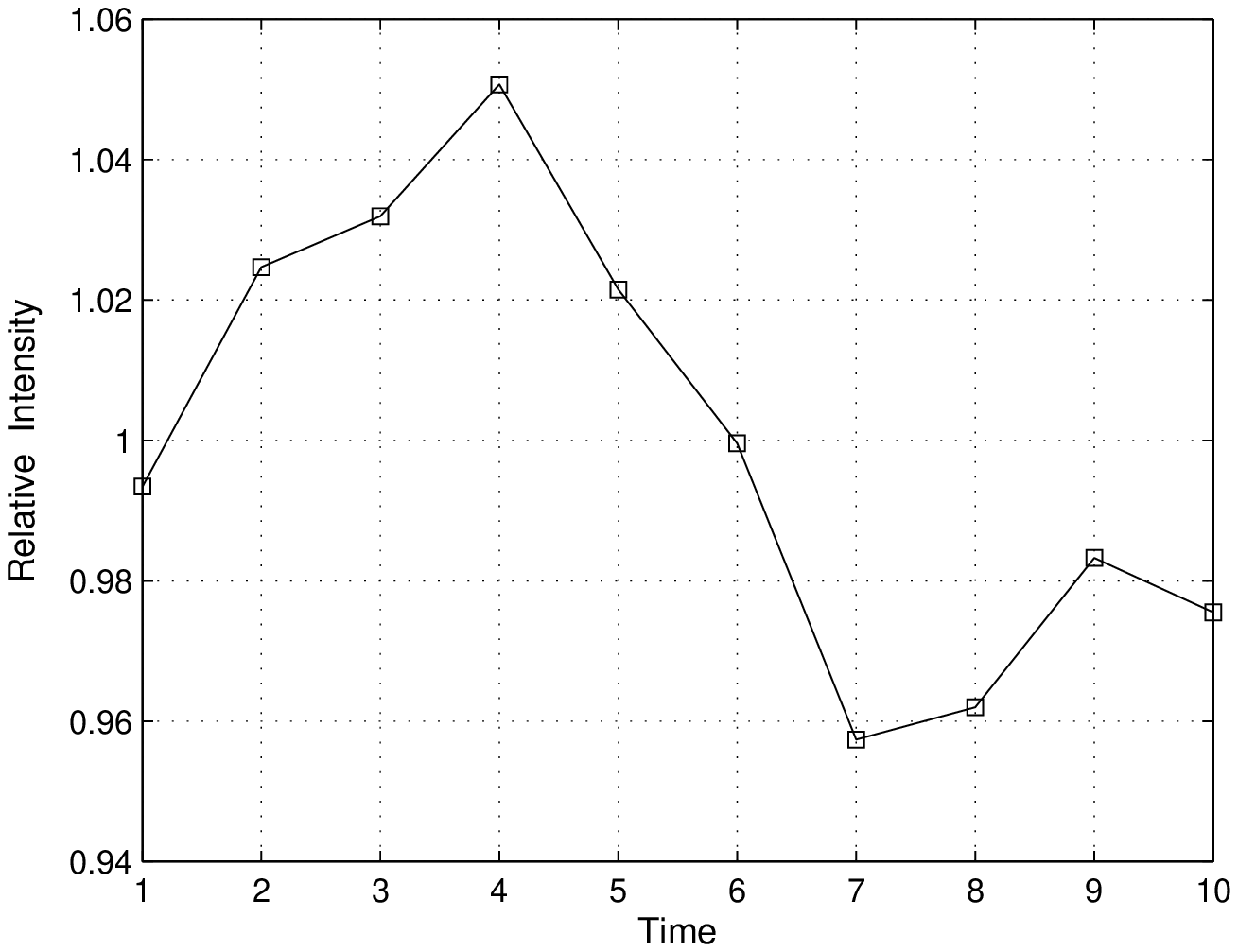,width = 1.05\linewidth} \caption{First object. Intensity variations.}\label{fig1}
\end{minipage}
\hfill
\begin{minipage}[t]{.45\linewidth} 
\centering
\epsfig{file = 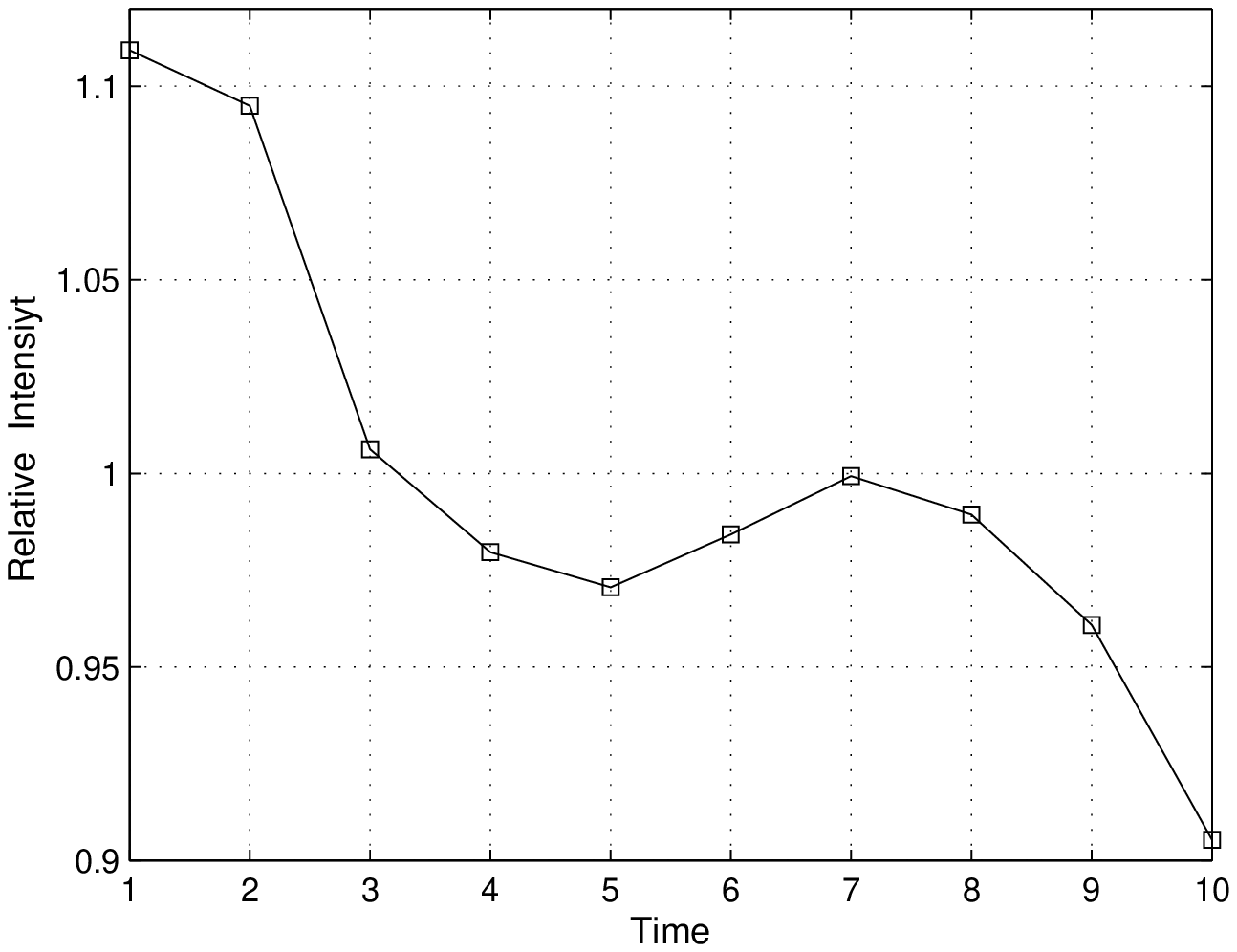,width = 1.0\linewidth} \caption{Second object. Intensity variations.}\label{fig2}
\end{minipage}
\end{figure}

\begin{figure}[!h]
\centering
\begin{minipage}[t]{.45\linewidth}
\centering
\epsfig{file = 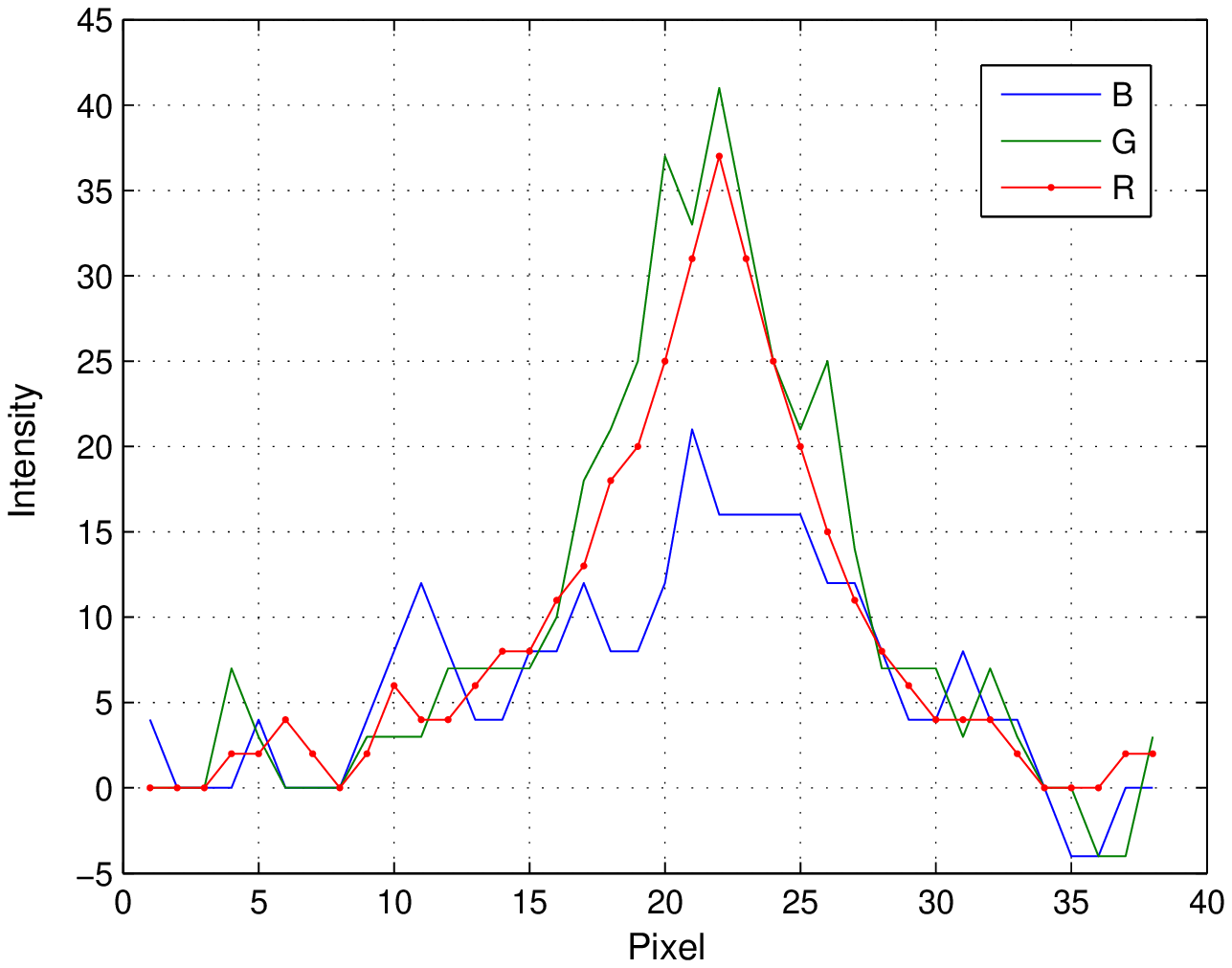,width = 1.05\linewidth} \caption{Second object. Color map.}\label{fig1}
\end{minipage}
\hfill
\begin{minipage}[t]{.45\linewidth} 
\centering
\epsfig{file = 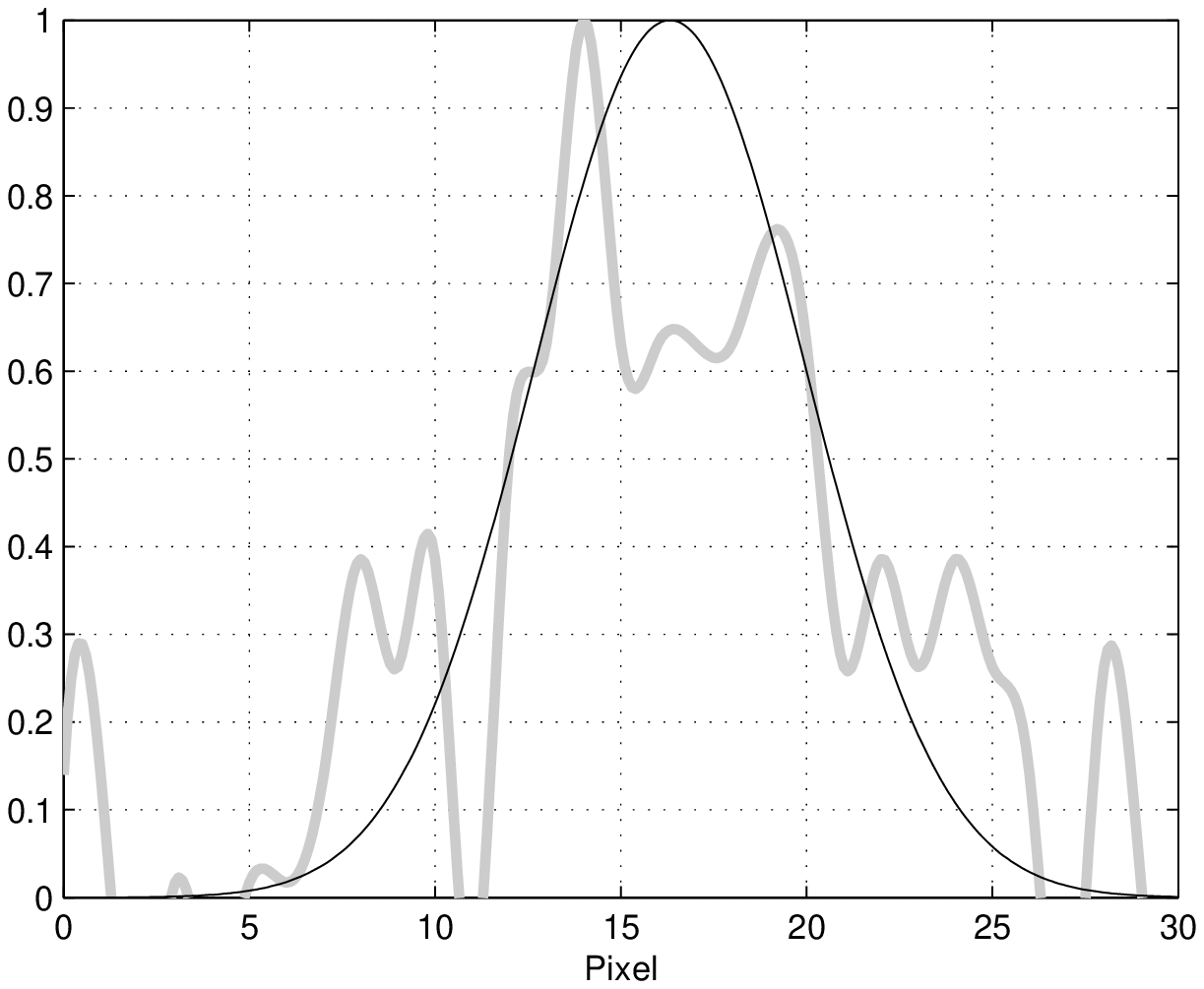,width = 1.0\linewidth} \caption{Second object. Size estimate. Gaussian approximation. FWHM = 3.2 arc min (1.3 km size). }\label{fig2}
\end{minipage}
\end{figure}

Observing UAPs, it becomes necessary to evaluate their characteristics. In particular, the sizes of bright objects can be determined if they shine with reflected sunlight. Practice shows that fireflies are visible only in the daytime sky. With the onset of twilight, their brightness decreases, and with the setting of the sun, they disappear from view. If their distance is known from parallax measurements, and if their albedo value is assumed, then one can easily determine the size of the object without angular resolution.

For calculations, we need to know the brightness of the daytime sky near the object and the distribution of energy in the spectrum of the Sun. The generic analytical expression of the spectrum of a clear daylight sky is given in \cite{Zagury}. The brightness of a clear blue sky for a zenith distance of 45 deg depending on the wavelength $ \lambda$ and the distribution of energy in the spectrum of the Sun are given in \cite{Allen}. In particular, a clear daylight sky brightness for $\lambda $ = 0.5 $\mu$m $F_{sky}$ = 4.5 $erg/(cm^{2}\cdot s \,\,  \cdot $\AA$ \cdot sr)$. The solar radiation flux $F_{sun}$ = 193 $erg/(cm^{2}\cdot s\,\, \cdot $\AA$)$.

We represent the light fluxes of the object and the sky background as:
\begin{equation}\label{}
I_{obj}=F_{sun} \cdot r^{2}/R^{2} \cdot \alpha
\end{equation}
\begin{equation}\label{}
I_{sky} = F_{sky}\cdot \Omega
\end{equation}
\begin{equation}\label{}
I_{obj} = \beta \cdot I_{sky}
\end{equation}
Here $r$ is the size of the object, $R$ is the distance to the object, $\alpha $ is the albedo, $\beta $ is the ratio of the brightness of the object to the brightness of the sky background, and $ \Omega$ is the solid angle of the pixel. This implies
\begin{equation}\label{}
r=R\cdot \sqrt{\beta\cdot F_{sky}\cdot \Omega / (F_{sun}\cdot \alpha)}
\end{equation}
Assuming $\beta $ = 1 and R = 200, 1400 and 2600 km, we obtain the following values for the object size in meters depending on the albedo in Table below.


\begin{center}
\begin{tabular}{ |c|c|c|c|c| } 
\hline
\multicolumn{5}{|c|}{albedo} \\
\hline
\multicolumn{5}{|c|}{Dimension, meters} \\
\hline
Distance & 0.2 & 0.1 & 0.01 & 0.001 \\
\hline
\multirow{3}{4em}{200 km 1400 km 2600 km} &8 &12&38&119\\ 
&59& 83&263& 832\\
&109&155&489&1545\\
\hline
\end{tabular}
\end{center}

Size estimates seem fantastic. The angular size observed gives a geometrical dimension about of 1 km (Fig. 9). FWHM of PSF (point spread function) gives about 3 pixels, i.e. about 1 arc minute. The object shows FWHM at about 3 arc minutes. Motion blur caused by the rapid movement of an object gives some increase in size. However, the problem does not disappear. Serious doubts are raised by the fact that these objects are not detected by radar, asteroid observers, and the military.
They are not seen at night. It follows that they shine by reflected sunlight.
Regular albedo of 0.2 (20\%) gives dimensions of hundreds of meters. It would seem that nothing prevents their easy detection.

An effective way to assume a very low albedo, which would make them undetectable to radar and the military.
Albedo less than 0.01 would seem to make them practically black bodies, not reflecting electromagnetic radiation. Indeed, a body smaller by an order of magnitude will emit two orders of magnitude less and will become invisible. 

Unfortunately, observations indicate large angular dimensions and, as a consequence, large geometric dimensions too (Fig. 9). It is possible to agree on large angular dimensions and theoretical estimates only by accepting extremely low albedo values, less than one per cent. This makes them invisible to the radar and the military. Not surprisingly, stealth technology makes aircraft undetectable to radar, but not to the eye.

It can be shown that objects with anomalously low albedo (less than 1 per cent) behave like a black body. The distance estimates given in APPENDIX give an accuracy of about 10 per cent. It can be assumed that in the troposphere they look like Phantoms, and outside the troposphere, they look like Cosmics.

\subsection*{\sc Color properties of bright flying objects}

Fig. 8 shows the color diagram of the object in the RGB Bayer filters. Object colors can be converted to the Johnson BVR astronomical color system using the color corrections published in \cite{Parka}. Semi-empirical relations are as follows:
\begin{equation}\label{}
(B-V)_{J}=1.47\cdot(B-G)+0.12
\end{equation}
\begin{equation}\label{}
(V-R)_{J}=(G-R)+0.23
\end{equation}
According to Fig. 8 the color index $(B-G)$ is 0.84. Hence $(B-V)_{J}$ is 1.35. Similarly $(V-R)_{J}$ is 0.23. Visually, such an object is perceived as very dark.

For control, we use the formulae (5, 6) to determine the color index $(B-V)_{J}$ in Mare Crisium of the Moon (Fig. 4). The calculated color index $(B-V)_{J}$ is equal 0.72. According to \cite{Thejll}, the measured color index $(B-V)_{J}$ is 0.75 $\pm $ 0.01.

Using the color chart in Fig. 8, we can restore the color image of the object in the Bayer RGB filters. We use a triple of row vectors [r g b] as indexes to specify the color. Note that the RGB intensity depends on the extinction correction known in \cite{Allen}: [0.91 0.82 0.73]. We have colotmap as $[rgb]=[1/0.91\,\, 1/0.82\,\, 0.43/0.73]\cdot albedo$. The restored images are shown in Fig. 10, 11.

\begin{figure}[!h]
\centering
\begin{minipage}[t]{.45\linewidth}
\centering
\epsfig{file = 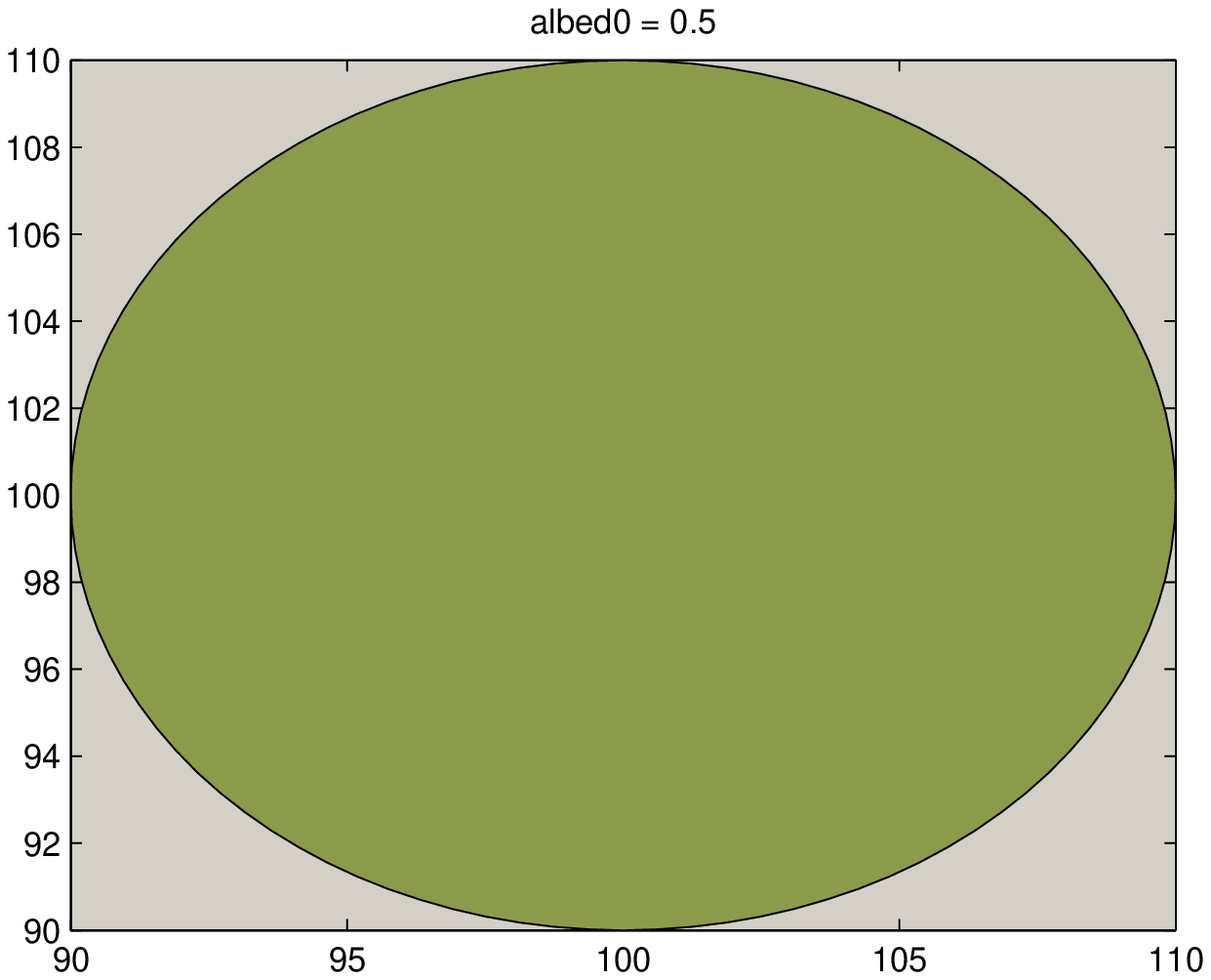,width = 0.85\linewidth} \caption{Object image. Albedo = 0.5.}\label{fig1}
\end{minipage}
\hfill
\begin{minipage}[t]{.45\linewidth} 
\centering
\epsfig{file = 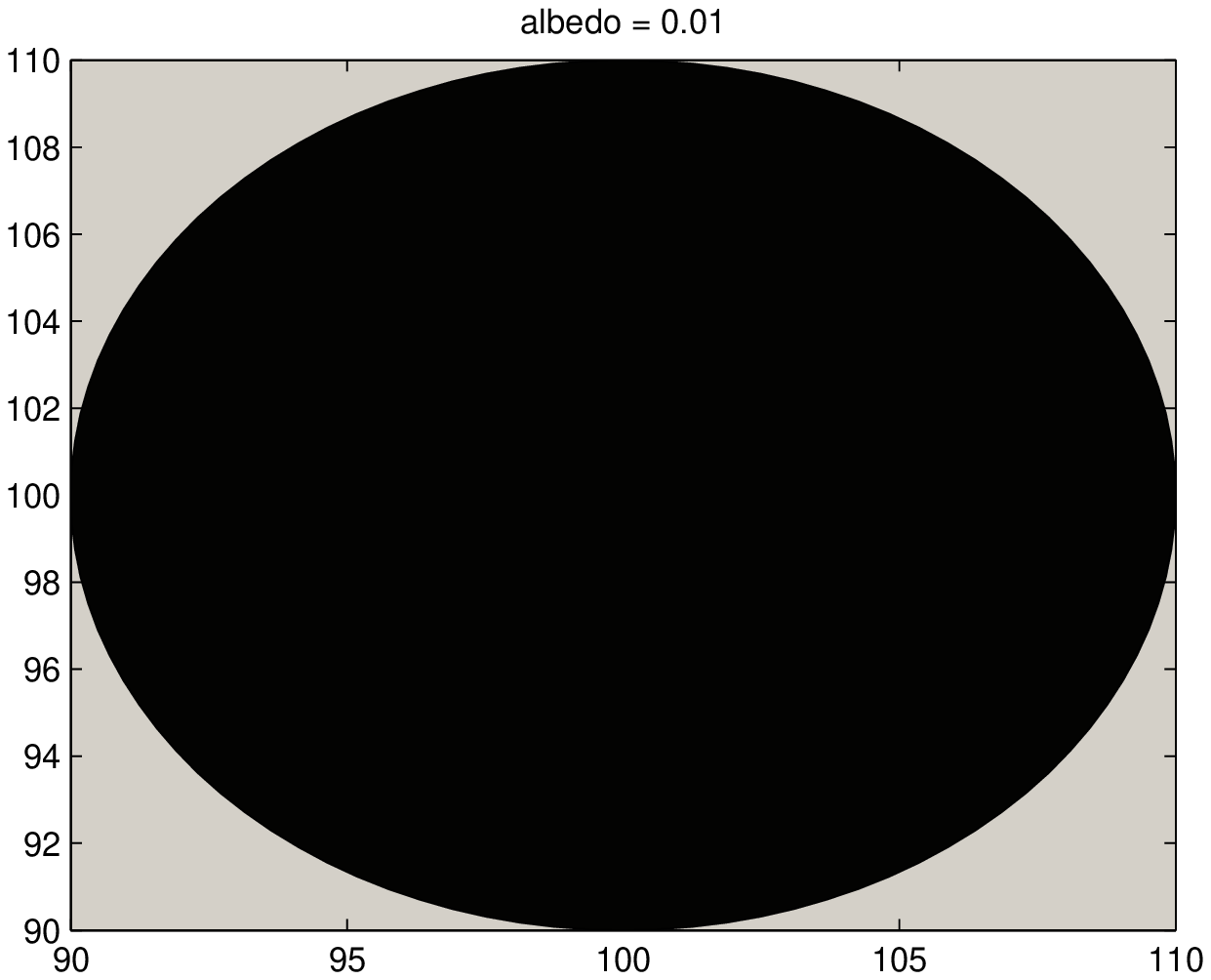,width = 0.85\linewidth} \caption{Object image. Albedo = 0.01. }\label{fig2}
\end{minipage}
\end{figure}

\section*{\sc Astronomical observations of dark flying objects}

Analysis shows that phantoms are absolutely black bodies, that is, they do not radiate and absorb the radiation incident on them. They become visible only when they are in the troposphere due to the fact that they partially screen the radiation of the atmosphere. The distance to them is determined by the image contrast value (see APPENDIX). Since the contrast is determined with an error due to the dynamic range of the camera and atmospheric noise, the detection of the phantom is limited to a distance of 10 - 14 km. In particular, this can explain the sudden appearance and disappearance of the phantom.

\subsection*{\sc Object 9252}

Object 9252 (Fig. 12, 13) was detected during the sky monitoring on August 24, 2018, at about 9 am(Figs. 12, 13). The clear sky was monitored in video mode with the Canon 5D Mark III camera in the Goloseevo district of Kyiv. Note that at 11:18 - 11:38 am a military equipment parade was held in connection with the anniversary of Ukraine's Independence.
The processing of observational data made it possible to establish the following characteristics of object 9252:

\begin{itemize}
    \item The moving object was observed for 0.17 seconds.
    \item At the beginning of the observed trajectory, the object was at a distance of 8.5 $\pm$ 2.0 km, at an altitude of 9.6 km.
    \item At the end of the observed trajectory, the object was at a distance of 13.3 $\pm$ 1.2 km, at an altitude of about 13.6 km and had angular dimensions comparable to the Moon.
    \item The object moved at a speed of about 96 degree/sec.
    \item The size of the thing is estimated to be about 94 meters.
    \item The object was shot in the falling sunlight, but it looked like a dark spot on the background of a bright sunny day sky. The contrast with the sky background in RGB colors at different distances of the object varied from 5 to 40\%.
\end{itemize}  

\subsection*{\sc Observations}

A multi-color DSLR camera with a full-scale sensor allows shooting in the field of view 90 degrees at a rate of 30 frames per second. The pictures 12, 13 show the object tracking. The beginning of the track is a double rectangle at the time of observations on August 24, 2018, at about 9 am. The length of the track corresponds to seventeen-hundredths of a second of time.

\begin{figure}[!h]
\centering
\begin{minipage}[t]{.45\linewidth}
\centering
\epsfig{file = 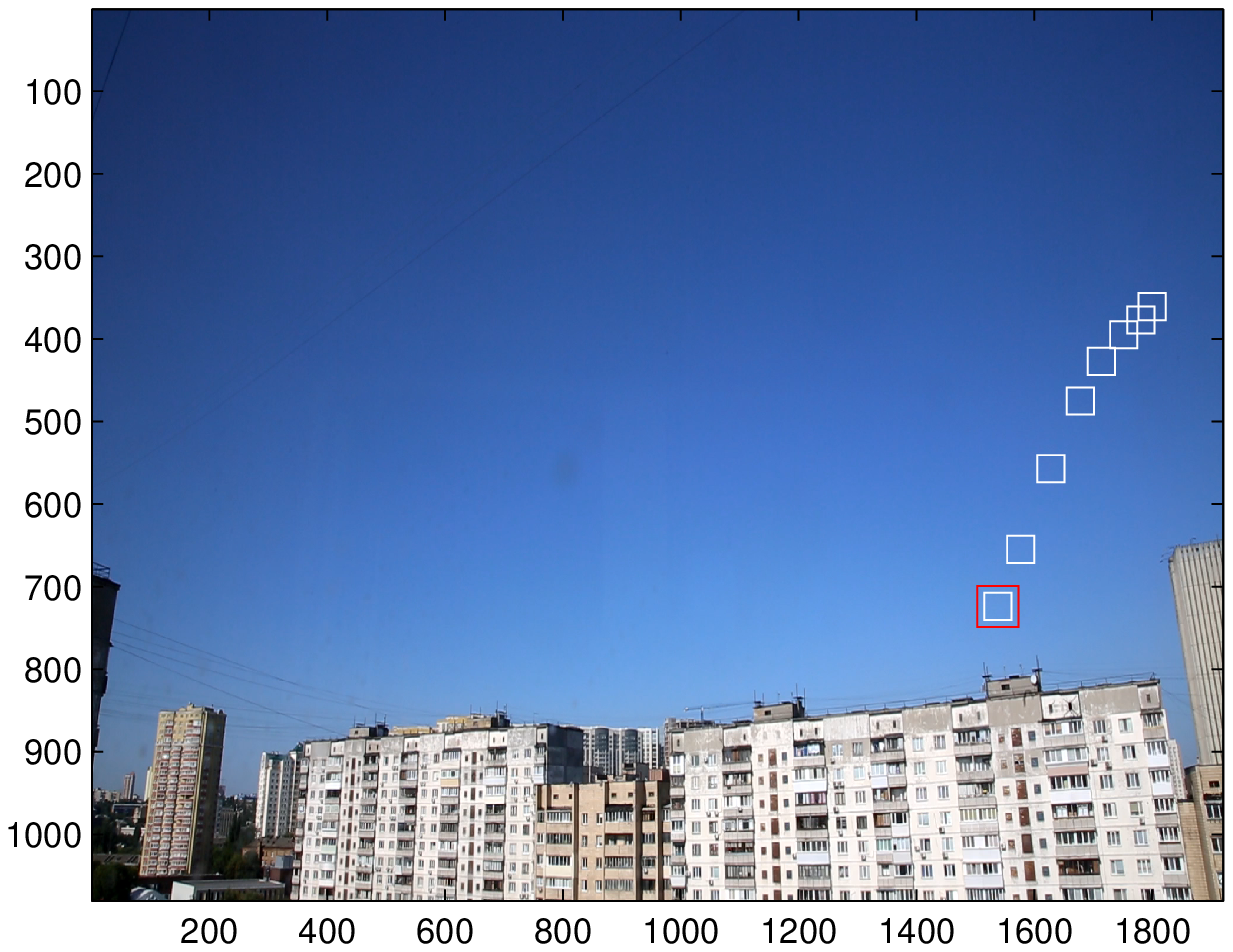,width = 0.9\linewidth} \caption{The picture shows the object tracking.}\label{fig1}
\end{minipage}
\hfill
\begin{minipage}[t]{.45\linewidth} 
\centering
\epsfig{file = 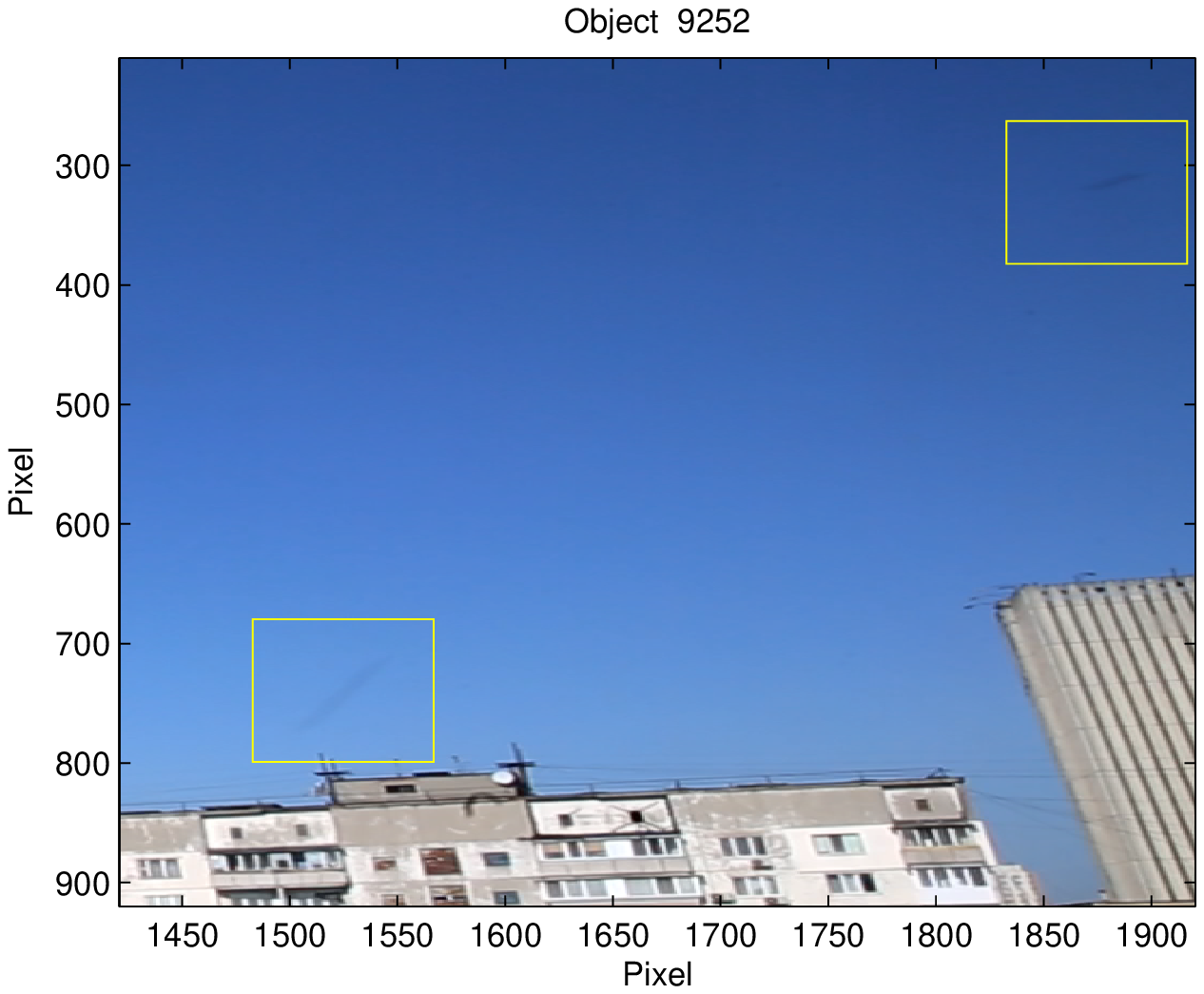,width = 0.9\linewidth} \caption{The picture shows the object tracking.}\label{fig2}
\end{minipage}
\end{figure}

\subsection*{\sc Technology of detection: blinking}

Fig. 14 shows the screen for detecting events using the method of blinking with a sampling frequency of 30 frames per second. The record length is about 8 minutes. The arrow marks event 9252, by the number of the frame in which it appeared. Other events are related to the flight of birds. The last peak is related to the plane taking off at the airport "In Kyiv".

The trajectory of the object is shown in pseudo colors (Fig. 15). For a short visibility time (0.17 s), the object increased the track size by twice. The width of the object track can reach up to 30 arc minutes, which is comparable to the size of the Moon. The angular velocity of the object is estimated at 96 degrees per second.

\begin{figure}[!h]
\centering
\begin{minipage}[t]{.45\linewidth}
\centering
\epsfig{file = 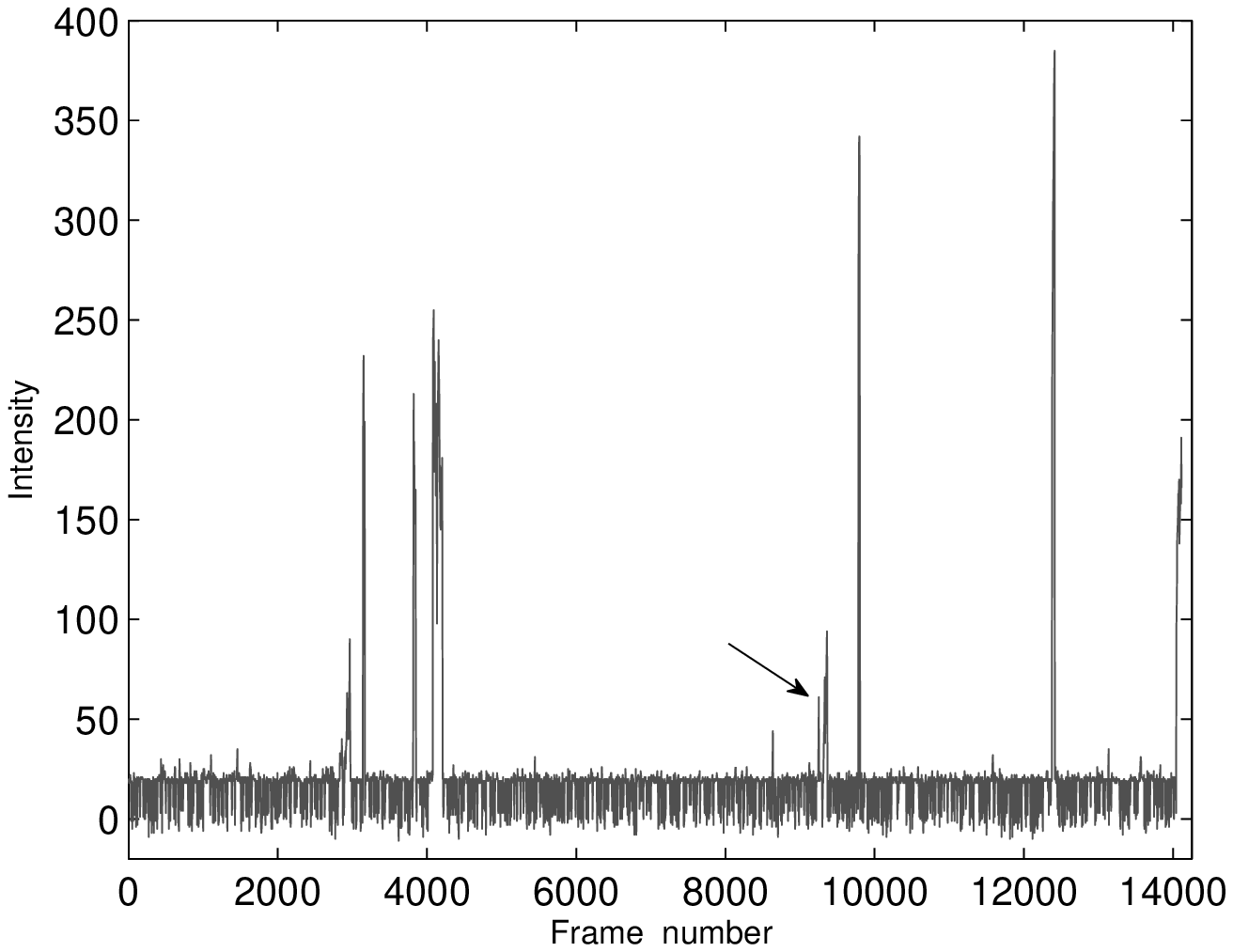,width = 0.9\linewidth} \caption{The method of blinking with a sampling frequency of 30
frames per second.}\label{fig1}
\end{minipage}
\hfill
\begin{minipage}[t]{.45\linewidth} 
\centering
\epsfig{file = 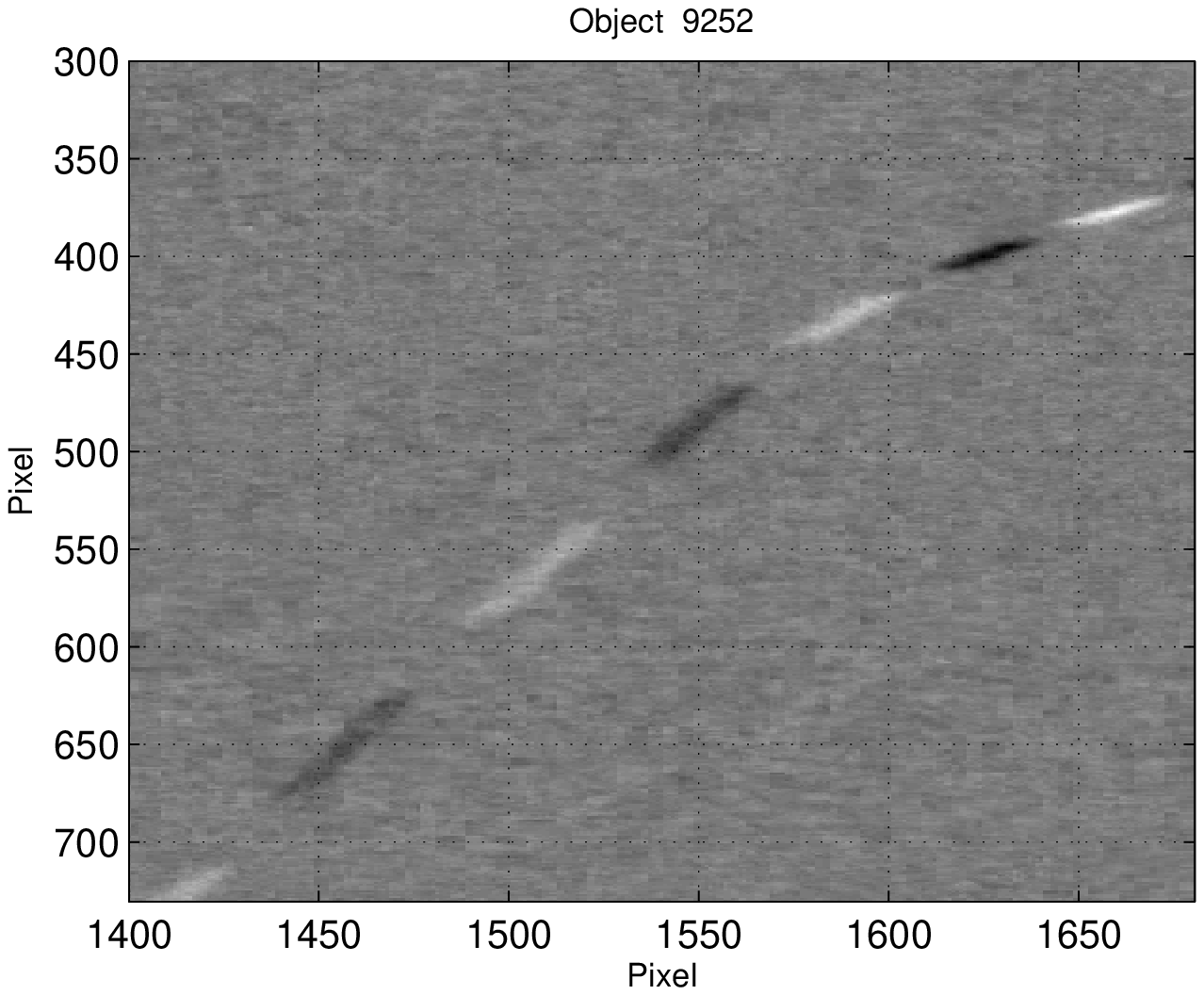,width = 0.9\linewidth} \caption{The trajectory of the object is shown in pseudo colors.}\label{fig2}
\end{minipage}
\end{figure}

\subsection*{\sc Restoring a distorted image of an object}

To eliminate the image motion blur caused by the rapid movement of the object, we construct a distorting point spread function for the case of motion blur PSF(L,$\theta $), where L is the length, and $\theta $ is the angle of the image motion. Motion blur is the convolution of an image of an object with a distorting function. The operation inverse to convolution is called deconvolution. As a result, we get an undistorted image of the object (Fig. 16).
To restore a distorted image of an object, we use inverted filtering using the Wiener filter. The kernel (hardcore) of the restored image is an ellipse measuring 5x10 pixels. The uneven distribution of brightness is a consequence of digital filtering.

\begin{figure}[!h]
\centering
\begin{minipage}[t]{.45\linewidth}
\centering
\epsfig{file = 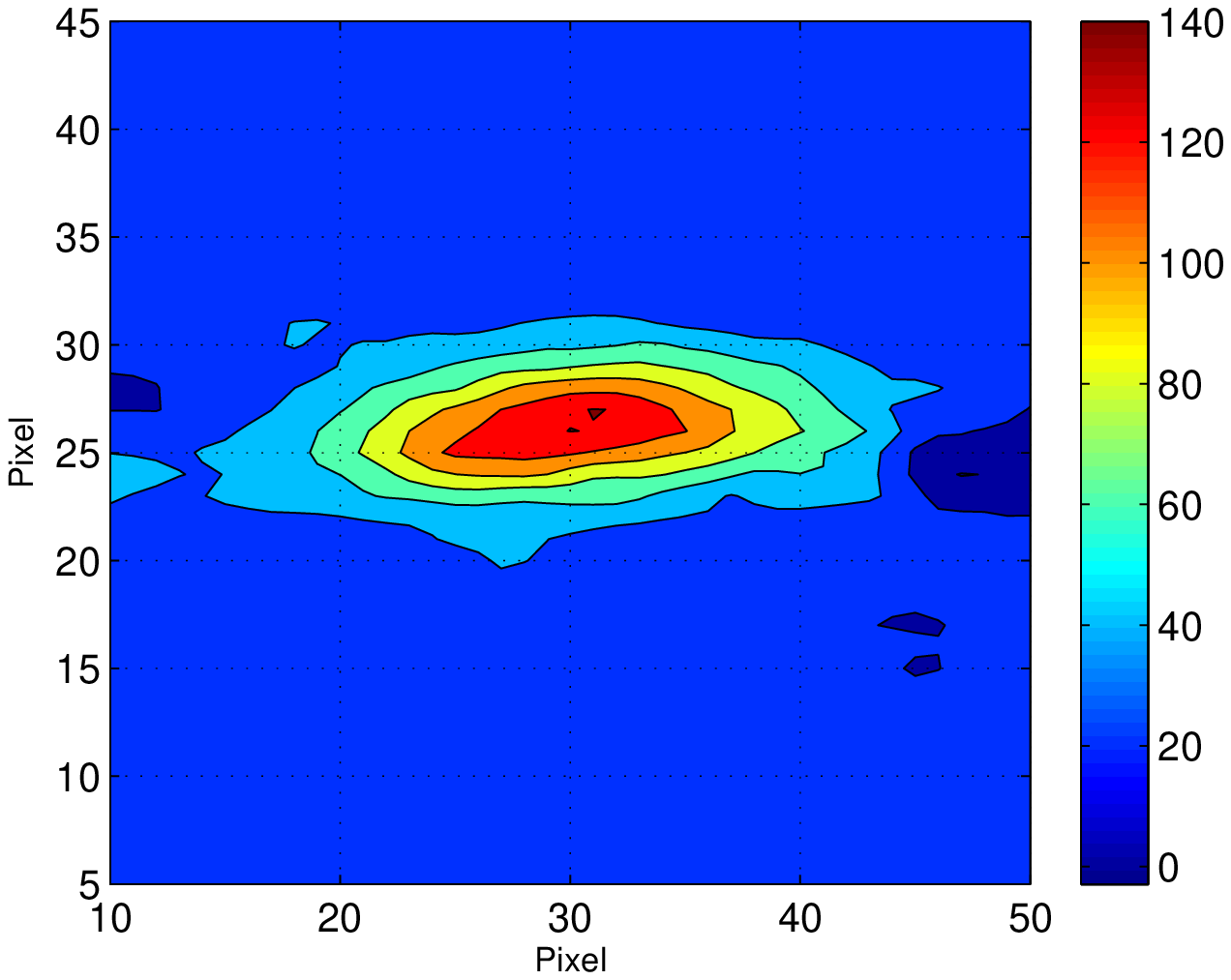,width = 0.9\linewidth} \caption{We get
an undistorted image of the object..}\label{fig1}
\end{minipage}
\hfill
\begin{minipage}[t]{.45\linewidth} 
\centering
\epsfig{file = 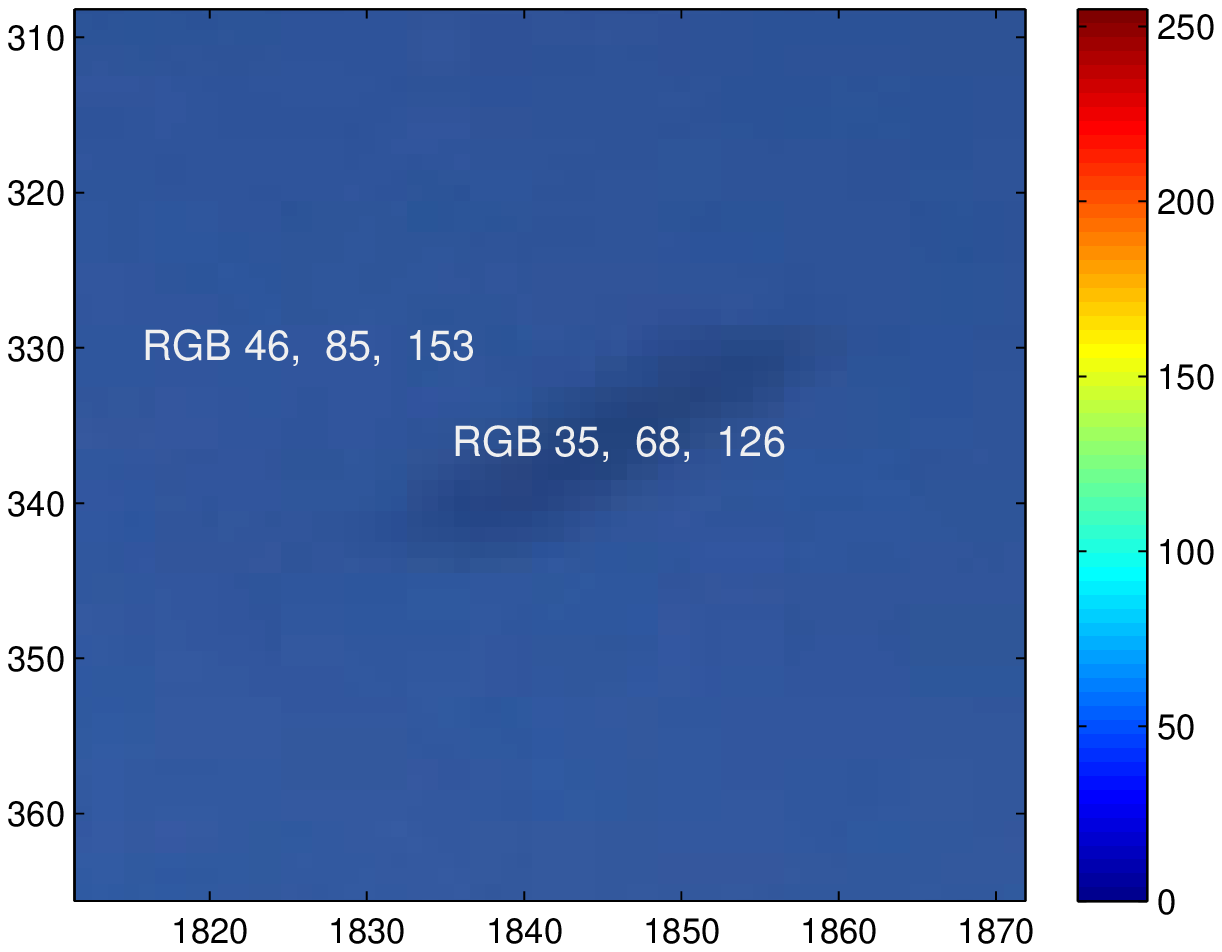,width = 0.9\linewidth} \caption{RGB data for the object and sky background.}\label{fig2}
\end{minipage}
\end{figure}

\subsection*{\sc Intrusion colourimetry}

We use a digital SLR camera with a full-scale CMOS sensor. The camera uses the RGB color standard. The RGB color space is easily converted to Johnson's astronomical BVR color system. This makes it possible to carry out the colourimetry of events in a quantitative form.
By measuring the brightness of the object and the sky background in 3 rays, one can estimate the magnitude of the event against the background of the daytime sky. The background of the daytime sky has a brightness from minus 3 to minus 5 magnitudes per square arc minute, depending on the distance to the center of the solar disk.

Fig. 17 represents the sample RGB data for the object and sky background at the end of the flight path. Similar data were obtained for the beginning of the trajectory. The data are shown in Table 1.

A comparison of the color indices of the background of the sky and the object confirms that they coincide within the error limits. From this, we can conclude that the object does not reflect solar radiation and does not have its own radiation. We observe the object only due to the fact that it partially shields the scattered light of the sky background. Thus, this serves as an important argument in favor of determining the distance by colourimetry. It also makes correct estimates of the range, height, and speed of the object obtained from the colorimetric data.

Let us make an important remark about the features of the object. The colors of the object and the background are the same (Fig. 17). But the object has less brightness compared to the sky background. The brightness of an object varies with distance. The object does not reflect or emit sunlight. In a sense, he is "invisible". It becomes visible due to the fact that it shields the scattered light of the sky behind it.


\begin{table}[h]
\caption{RGB data of object and background } 
\centering 

\begin{tabular}{|c|c|c|}
  \hline
  Time & Background, mag &  Object, mag \\
  \hline
  In the beginning & G-R = -0.47 $\pm$ 0.02  & G-R = -0.46 $\pm$ 0.04 \\
  & B-G = -0.42 $\pm$ 0.01 &  B-G = -0.4 $\pm$ 0.05\\
  \hline
  In the end & G-R = -0.66 $\pm$ 0.01 & G-R = -0.66 $\pm$ 0.08 \\
  & B-G = -0.64 $\pm$ 0.01 & B-G = -0.69 $\pm$ 0.02 \\
  \hline
\end{tabular}
\end{table}

We have shown that object 9252 does not reflect solar radiation (the albedo is zero) and, therefore, formula (9) is rigorous both for estimating the distance to the object and for estimating the height and speed of the object.

Figs. 18, 19 shows photometric sections of the trace in the RGB rays at the beginning and end of the trajectory. The extinction in the object is normalized to the intensity of the sky background in the vicinity of the object. The first conclusion is that the residual intensity in all filters is the same. This means that the colors of the object and the background are the same, that the object does not have its own radiation and, most importantly, does not reflect solar radiation. It is theoretically possible that it reflects equally in all wavelengths, i.e. the object is a completely grey body, which is apparently unlikely.
The second conclusion is that the residual intensity in all filters is different at the beginning and end of the trajectory. This means that the air mass separating the object from the observer is different. Moreover, it is directly proportional to the distance to the object. In fact, this provides a way to measure distance using colourimetry methods.
The third conclusion is that the trace width is different at the beginning and end of the path and is 14 pixels at the beginning and 9 - 10 pixels at the end. As will be seen below, this is fully consistent with the change in distances to the object.

\begin{figure}[!h]
\centering
\begin{minipage}[t]{.45\linewidth}
\centering
\epsfig{file = 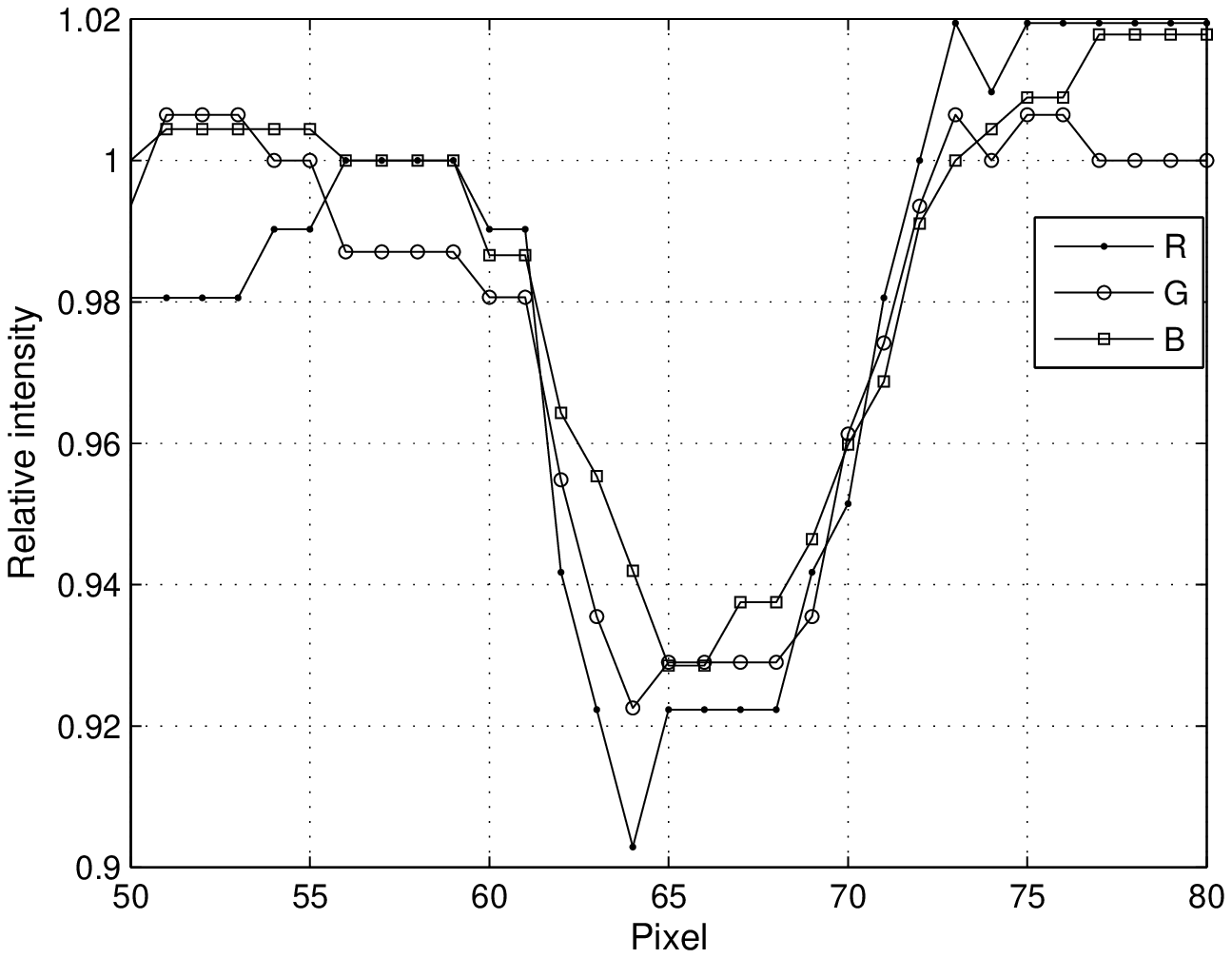,width = 0.9\linewidth} \caption{Beginning. Size 14 pxs.}\label{fig1}
\end{minipage}
\hfill
\begin{minipage}[t]{.45\linewidth} 
\centering
\epsfig{file = 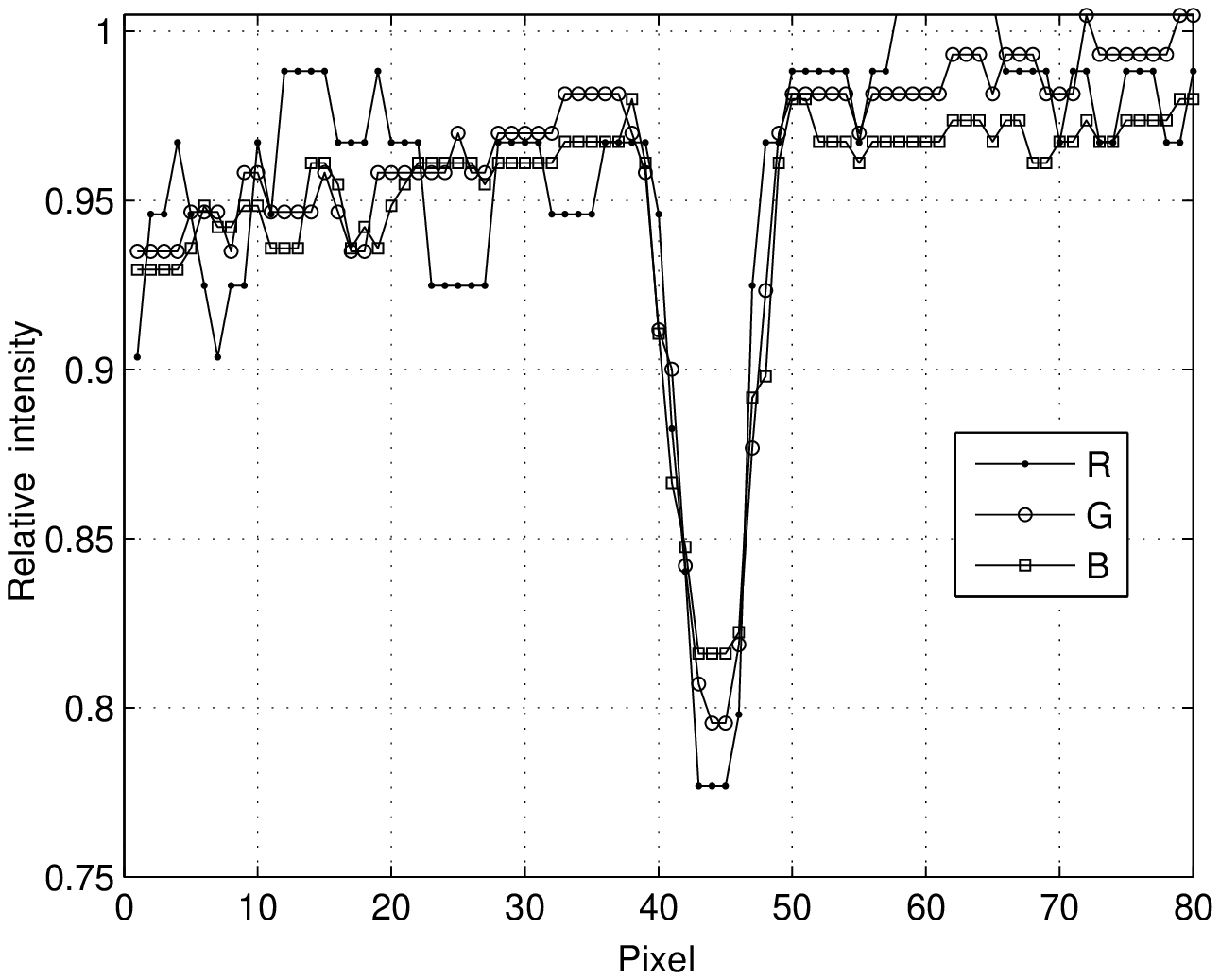,width = 0.9\linewidth} \caption{End. Size 10 pxs.}\label{fig2}
\end{minipage}
\end{figure}

Figures 18 and 19 provide estimates of the distances to the object according to formulae in APPENDIX in the approximation of a homogeneous atmosphere with the object's height above the horizon and residual intensities of 0.92 and 0.80. The following estimates of the distance, height and size of the thing were obtained:
\begin{itemize}
    \item Distance 8.5 km, height 9.6 km at the beginning  of the trajectory.
    \item Distance 13.3 km, height 13.6 km at the end of the trajectory.
    \item The size is 94.0 and 93.5 meters (14 and 9 pixels) at the beginning and end of the trajectory.
    \item The length of the trajectory is 5.4 km.
    \item The speed of the object on the trajectory is 31.8 km/sec.
\end{itemize}

\subsection*{\sc Some considerations regarding birds and insects}

At a distance of 1 km with an angular resolution of 1 arc minute, the linear size of the object is 50 cm. Let the characteristic size of the birds be 20 and 10 cm. Then the birds at a distance of 400 and 200 m will look like point objects. At smaller distances, the birds will look like structured objects. Assuming the speed of the birds is 10 m/s, it is easy to estimate their angular velocity of about one and two degrees per second.

From this, it is easy to conclude that birds cannot be phantoms.

Insects are the primary source of interference in UAP observations. They partly shine with reflected solar radiation (in proportion to their albedo) and shield atmospheric radiation.
The intensity of their radiation is proportional to $r^{2}/R^{2}$, where $r$ is the characteristic size and $R$ is the distance to the object. As the object moves away, the intensity drops. In practice, insects become invisible at distances greater than 15 m due to competition with atmospheric radiation intensity and measurement errors.
Selection of insects can be carried out according to several factors: (1) they move along curvilinear trajectories; (2) according to the color indices of their radiation; (3) for observations from two points.



\subsection*{\sc ARRIVAL 2}

\begin{itemize}
    \item An object appears suddenly at a distance of 11 km and an altitude of about 2 km (Fig. 20, 21).
    \item Approached at a distance of about 4 km.
    \item The object turned around and left. The evolution of the shape of the object in the process of movement is clearly visible (Fig. 24).
   \item The object was moving at a speed of about 10 km/s.
    \item An object at a distance of about 11 km looks like a faint spot against the background of the daytime sky. Clearly visible to the naked eye as a black object at a distance of about 4 km (Fig. 25, 26).
    \item Colorimetry of traces of an object allows us to determine its main characteristics: range, height, size, speed (Fig. 22, 23).
    \item Time of existence - 0.72 sec. Size 79 $\pm $ 5 m
    \item Altitudes: at the beginning 2.4 km, at the end 1.9 km
\end{itemize}

\begin{figure}[!h]
\centering
\begin{minipage}[t]{.45\linewidth}
\centering
\epsfig{file = 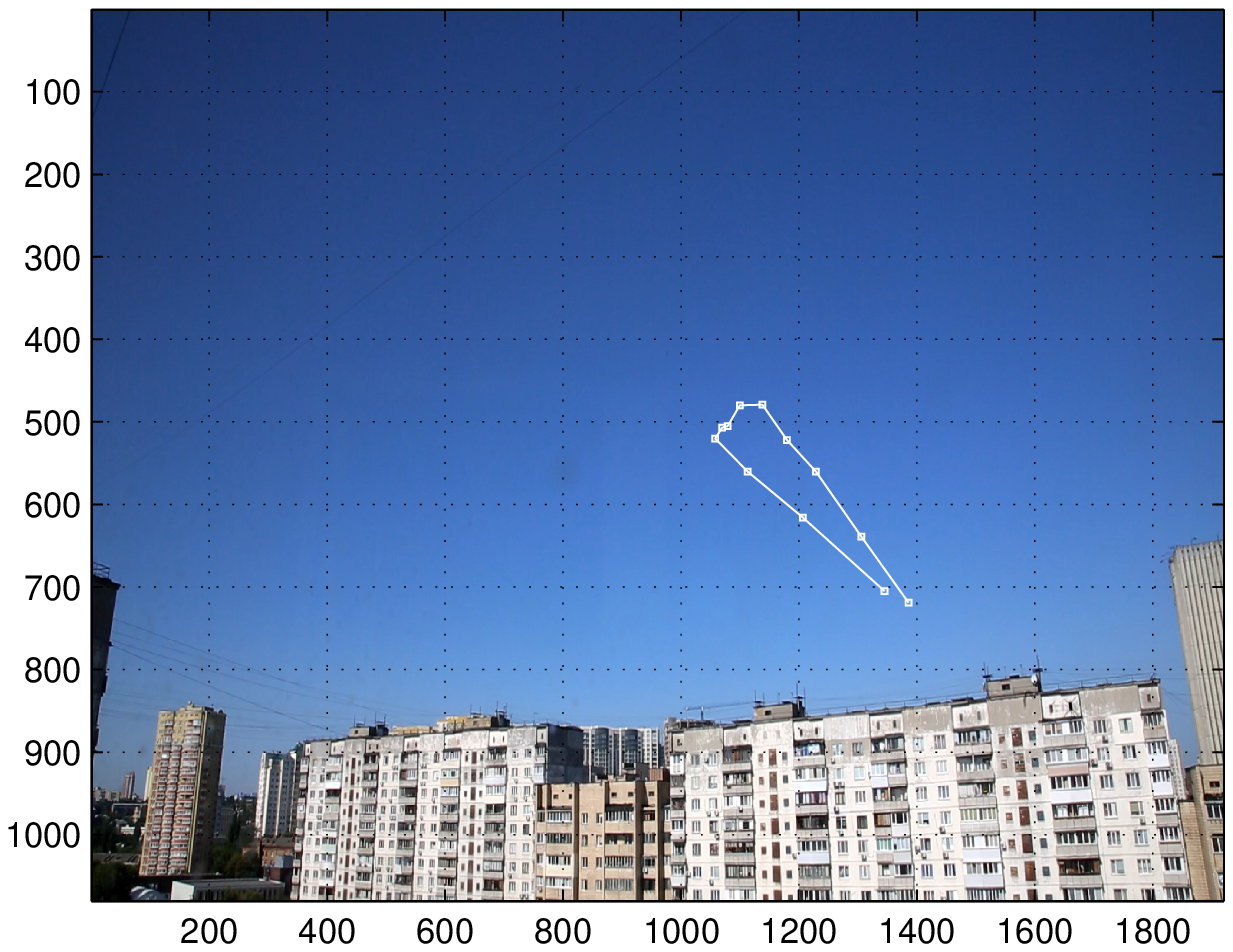,width = 0.9\linewidth} \caption{Arrival No. 2.}\label{fig1}
\end{minipage}
\hfill
\begin{minipage}[t]{.45\linewidth} 
\centering
\epsfig{file = 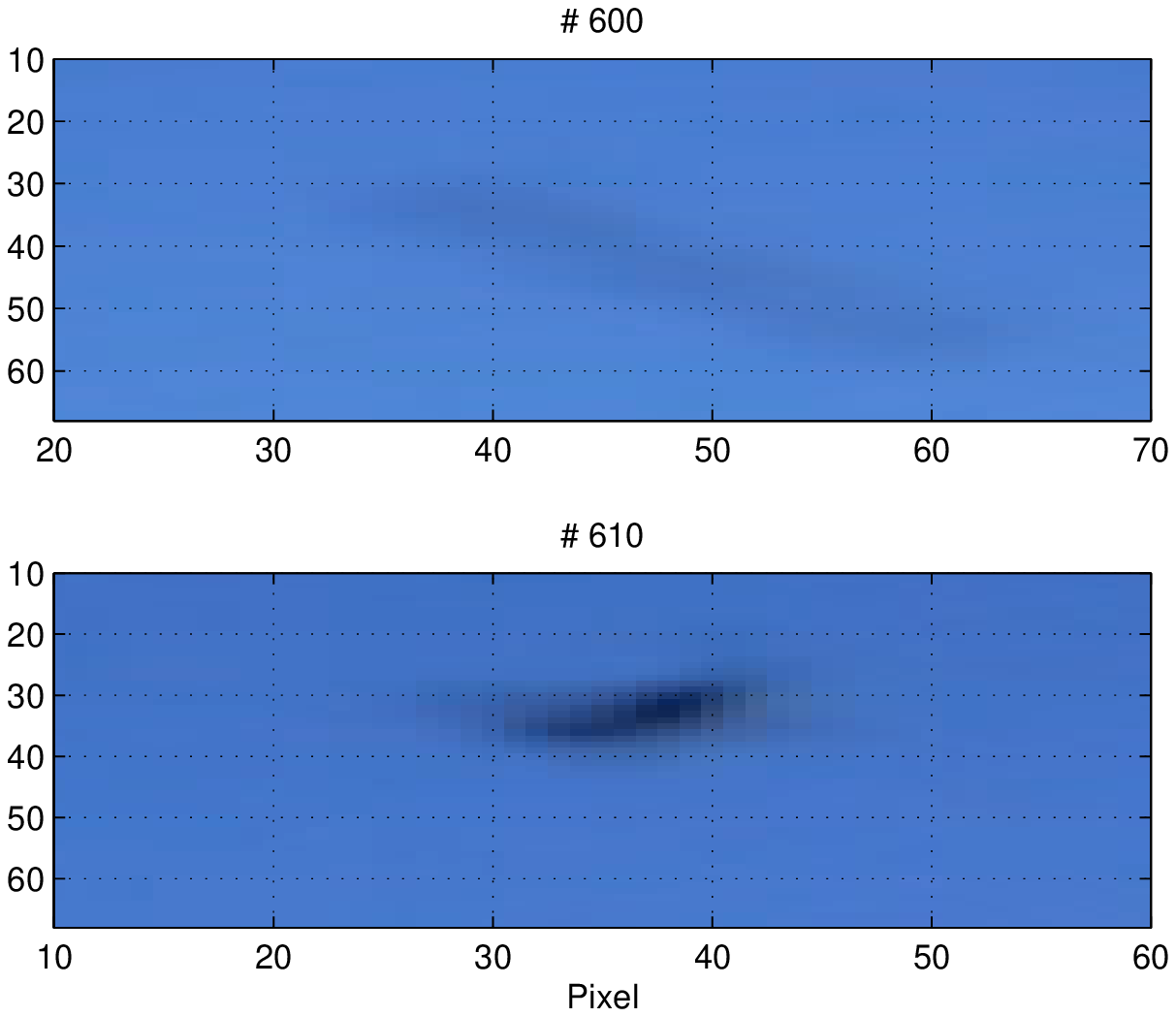,width = 0.9\linewidth} \caption{Distance 11 and 4 km.}\label{fig2}
\end{minipage}
\end{figure}

\begin{figure}[!h]
\centering
\begin{minipage}[t]{.45\linewidth}
\centering
\epsfig{file = 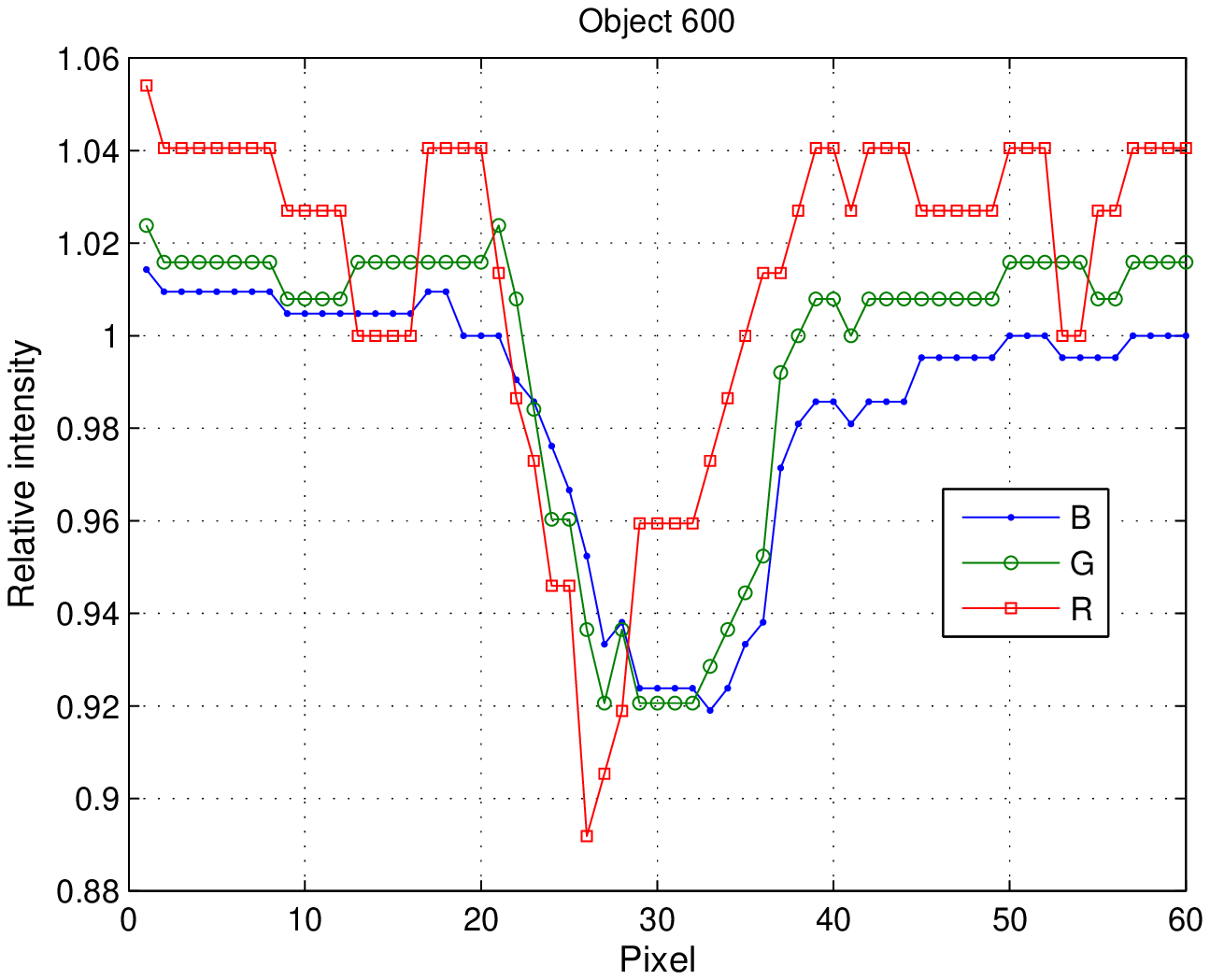,width = 0.9\linewidth} \caption{Distance 11 km..}\label{fig1}
\end{minipage}
\hfill
\begin{minipage}[t]{.45\linewidth} 
\centering
\epsfig{file = 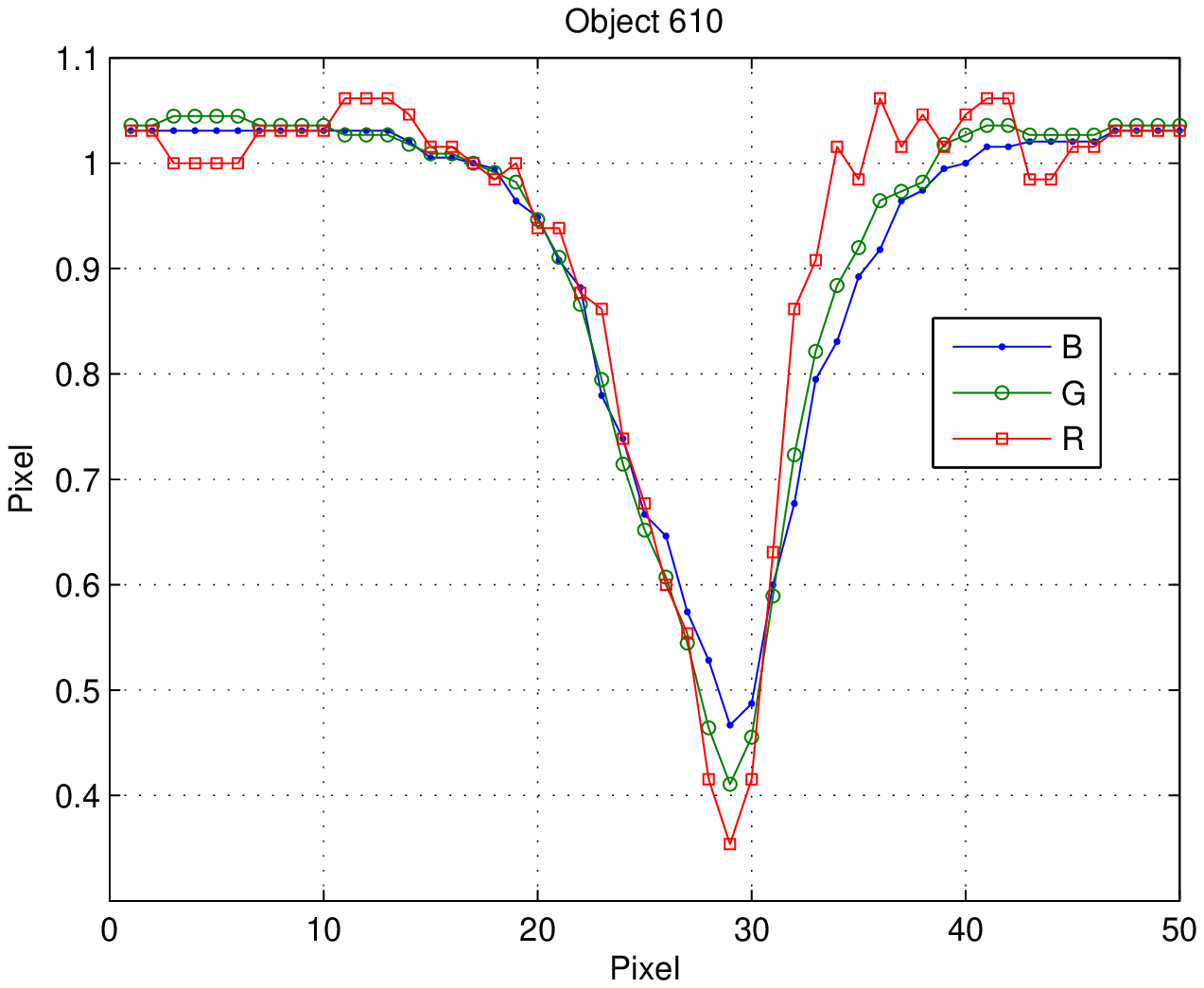,width = 0.9\linewidth} \caption{Distance 4 km.}\label{fig2}
\end{minipage}
\end{figure}

\begin{figure}[h]
\centering
\resizebox{0.75\hsize}{!}{\includegraphics[angle=000]{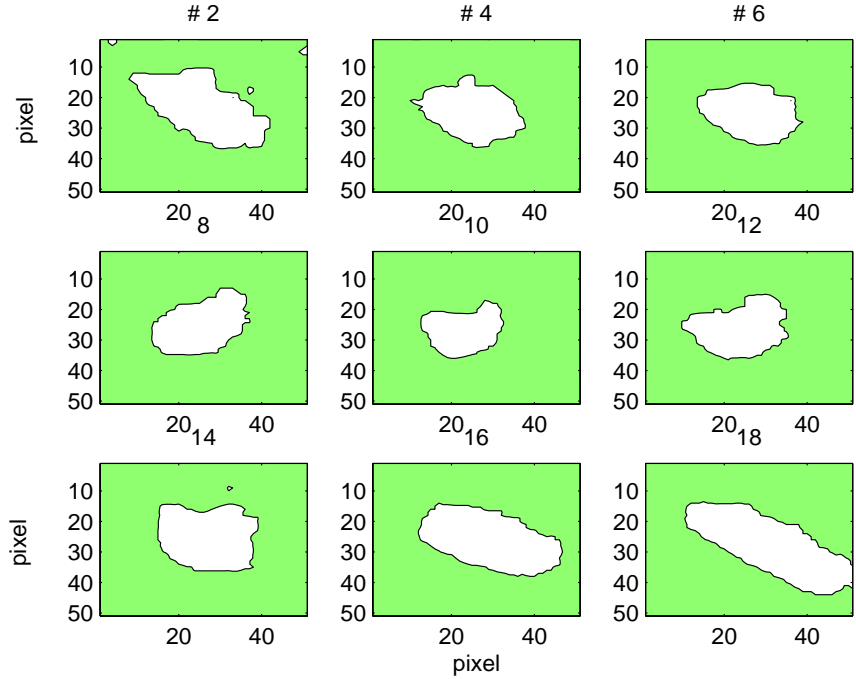}}%
\caption{The evolution of forms. Contour images of an object in motion.} \label{figure: BrightObj_16.eps}
\end{figure}

\begin{figure}[!h]
\centering
\begin{minipage}[t]{.45\linewidth}
\centering
\epsfig{file = 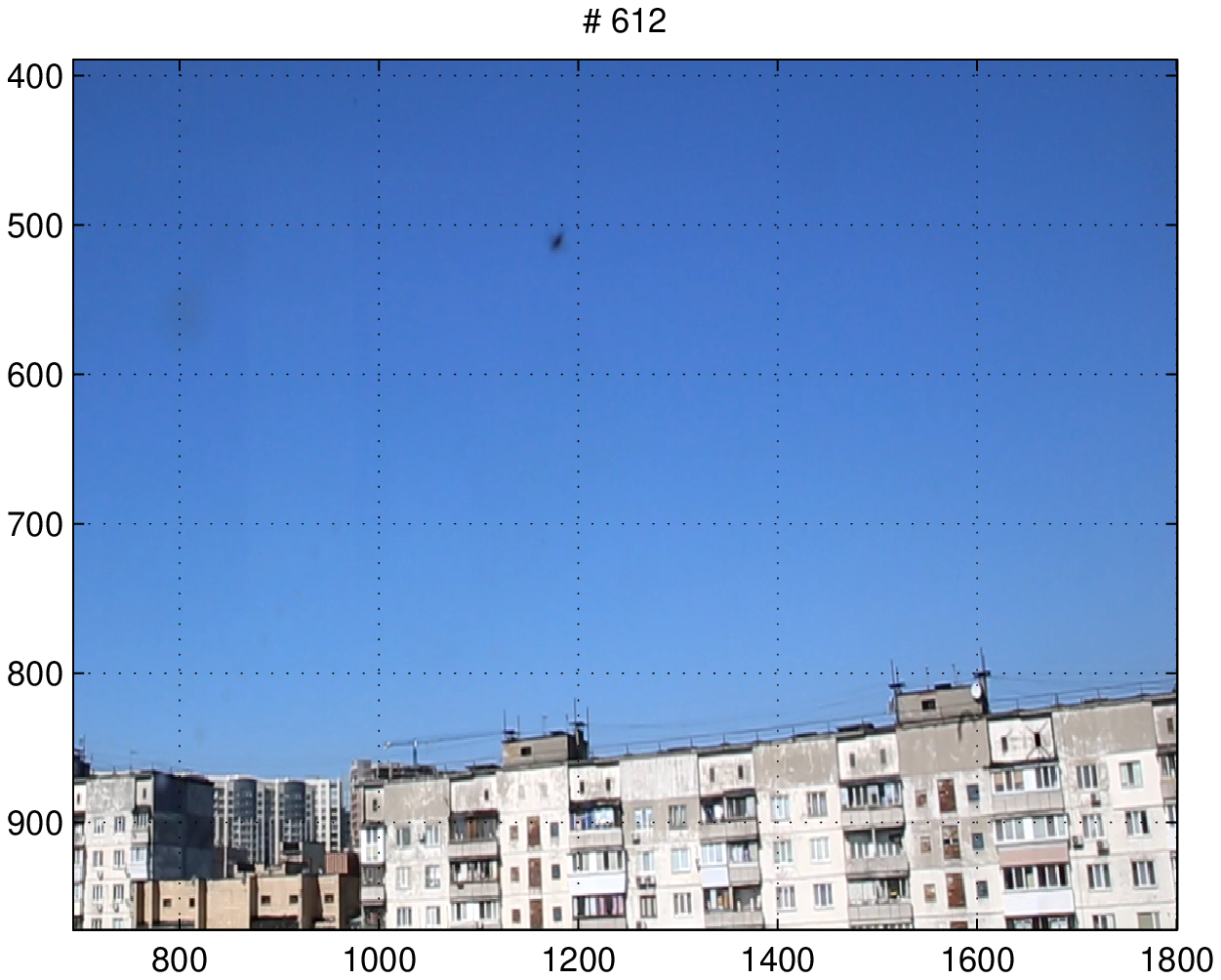,width = 0.9\linewidth} \caption{Phantom visible to the naked eye.}\label{fig1}
\end{minipage}
\hfill
\begin{minipage}[t]{.45\linewidth} 
\centering
\epsfig{file = 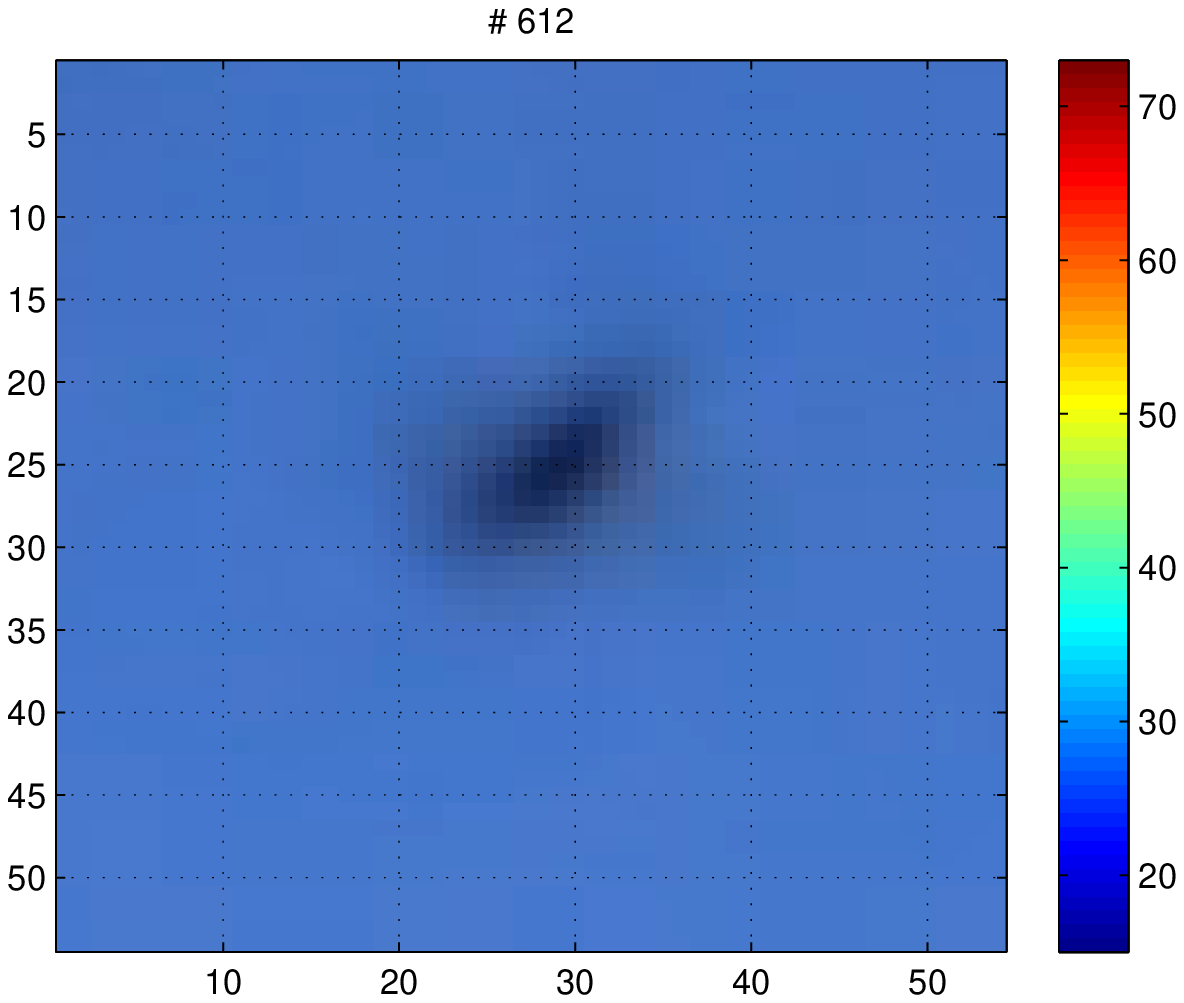,width = 0.9\linewidth} \caption{Phantom visible to the naked eye.}\label{fig2}
\end{minipage}
\end{figure}

\subsection*{\sc ARRIVAL 3}

\begin{itemize}
    \item Kyiv, August 24, 2018, 9 am
    \item Object's track  (Fig. 27).
    \item An object about 45 m in size descended from the stratosphere to a height of about 7 km to the Kyiv airport at a speed of about 30 km/sec. Then, in about 1 second, the object dropped from 7 to 1 km without approaching us.
    \item After that, the object rose and left. The whole episode took 2.2 seconds
    \item The object's colormap gives a distance of about 12 km  (Fig. 28).
\end{itemize}

\subsection*{\sc ARRIVAL 4}

\begin{itemize}
    \item Kyiv, September 26, 2018.
    \item Object's track  (Fig. 29).
    \item Distance 10 km, Height 4.5 km, Size 79 m. 
    \item The whole episode took 0.2 sec.
    \item An ellipsoidal shape object maneuver in 0.13 seconds (Fig. 30).
\end{itemize}    

\subsection*{\sc ARRIVAL 5}

\begin{itemize}
    \item Vinarivka, October 15, 2022.
    \item Object's track  (Fig. 31, 32).
    \item The whole episode took 0.128 sec.
    \item Object image (Fig., 33).
    \item The object's color maps give a distance of about 13.7 km (Fig. 32, 34).
    \item Speed is 2.3 km/sec = 7.2 Mach.
    \item Size is 20 meter.
\end{itemize}

\begin{figure}[!h]
\centering
\begin{minipage}[t]{.45\linewidth}
\centering
\epsfig{file = 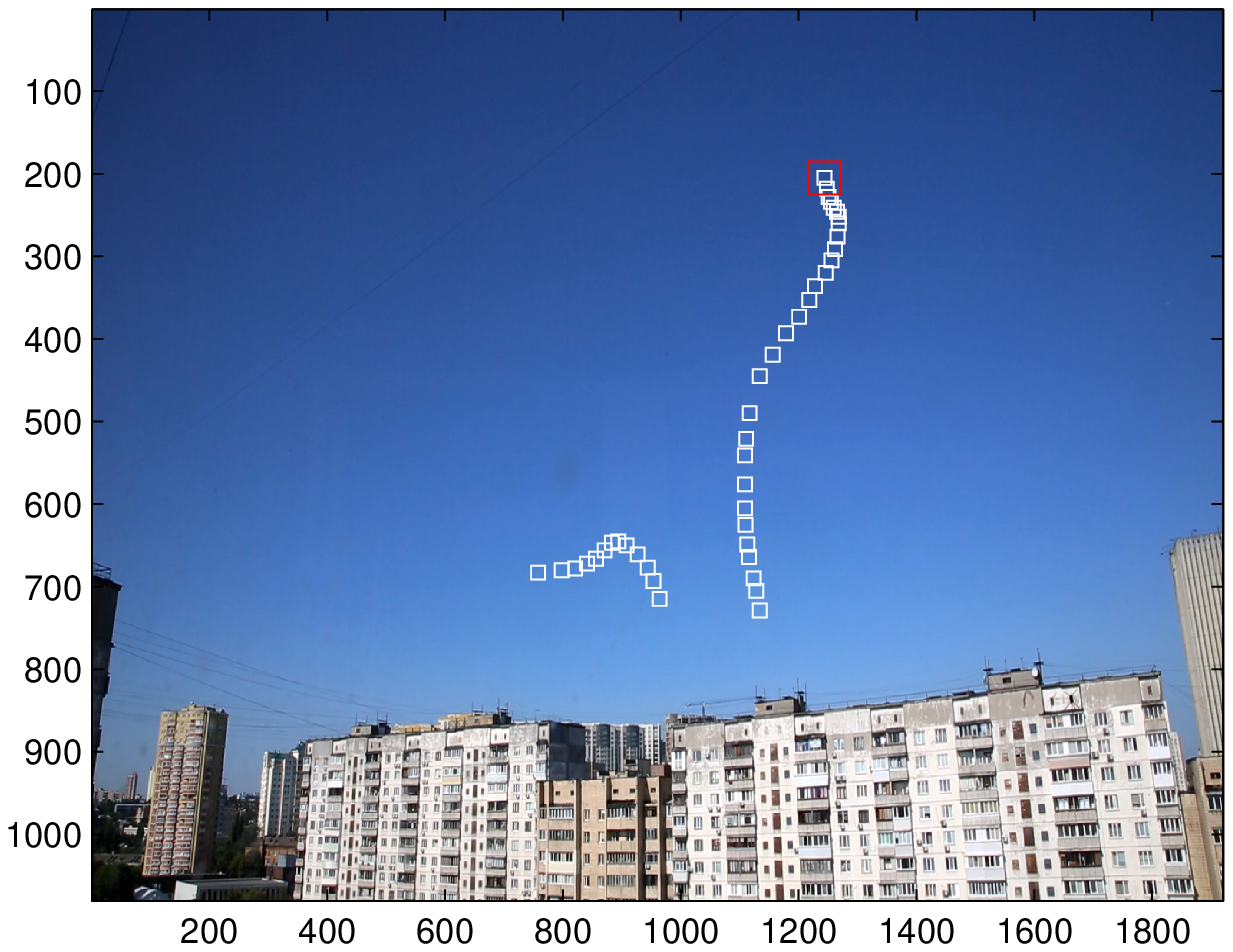,width = 0.9\linewidth} \caption{Arrival No. 3.}\label{fig1}
\end{minipage}
\hfill
\begin{minipage}[t]{.45\linewidth} 
\centering
\epsfig{file = 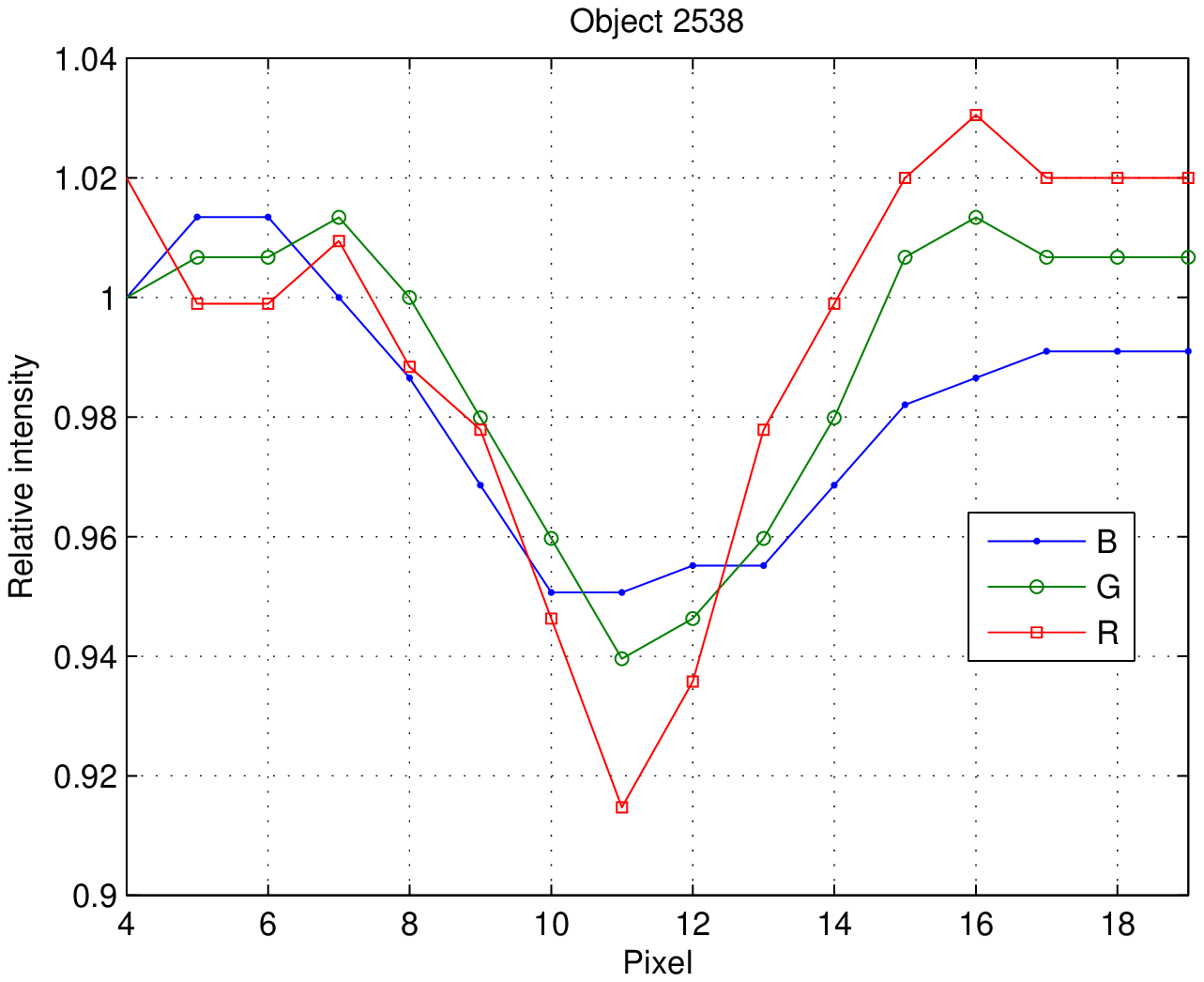,width = 0.9\linewidth} \caption{Arrival No. 3.}\label{fig2}
\end{minipage}
\end{figure}

\begin{figure}[!h]
\centering
\begin{minipage}[t]{.45\linewidth}
\centering
\epsfig{file = 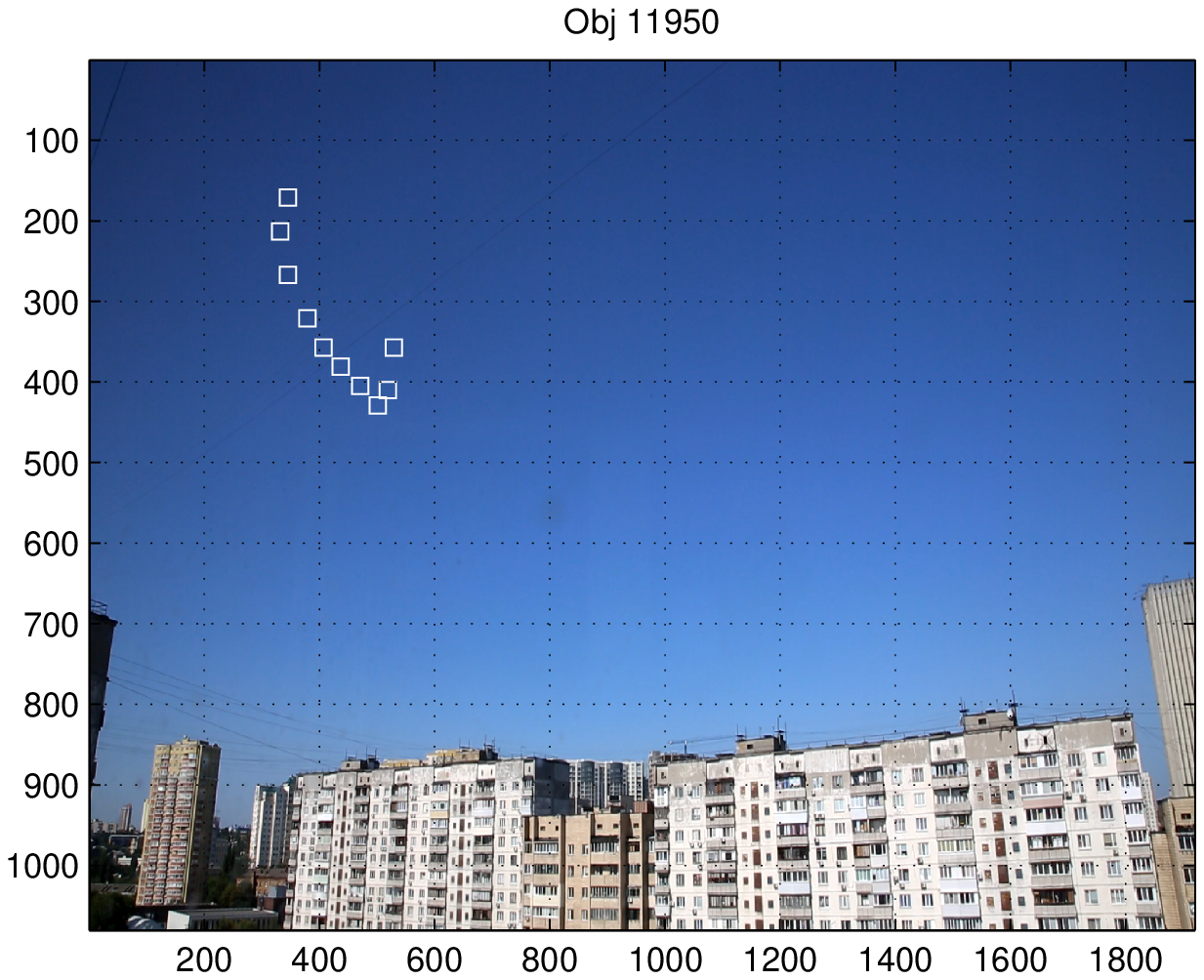,width = 0.9\linewidth} \caption{Arrival No. 4.}\label{fig1}
\end{minipage}
\hfill
\begin{minipage}[t]{.45\linewidth} 
\centering
\epsfig{file = 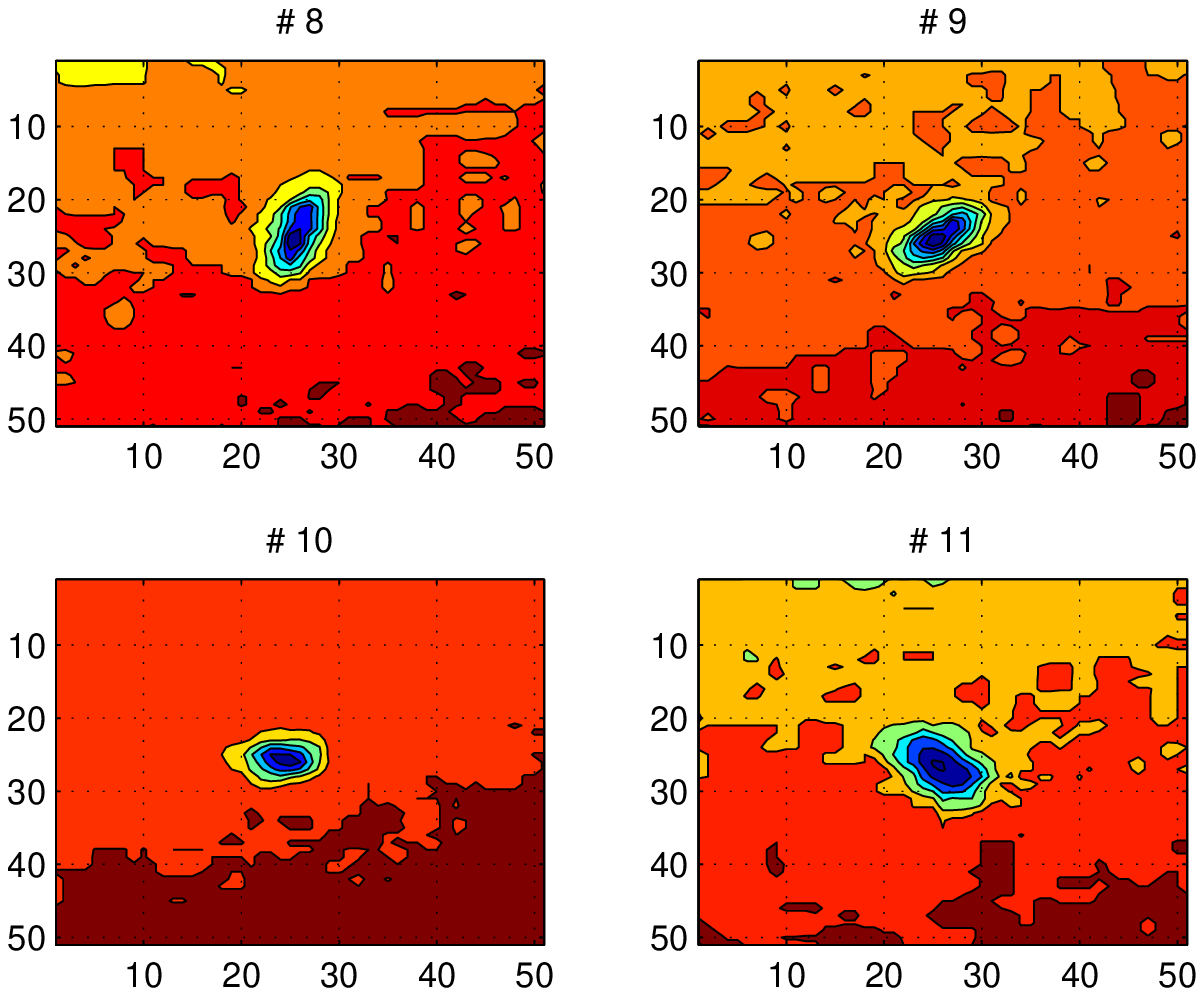,width = 0.9\linewidth} \caption{Object manoeuvres in 0.13 sec.}\label{fig2}
\end{minipage}
\end{figure}

\begin{figure}[!h]
\centering
\begin{minipage}[t]{.45\linewidth}
\centering
\epsfig{file = 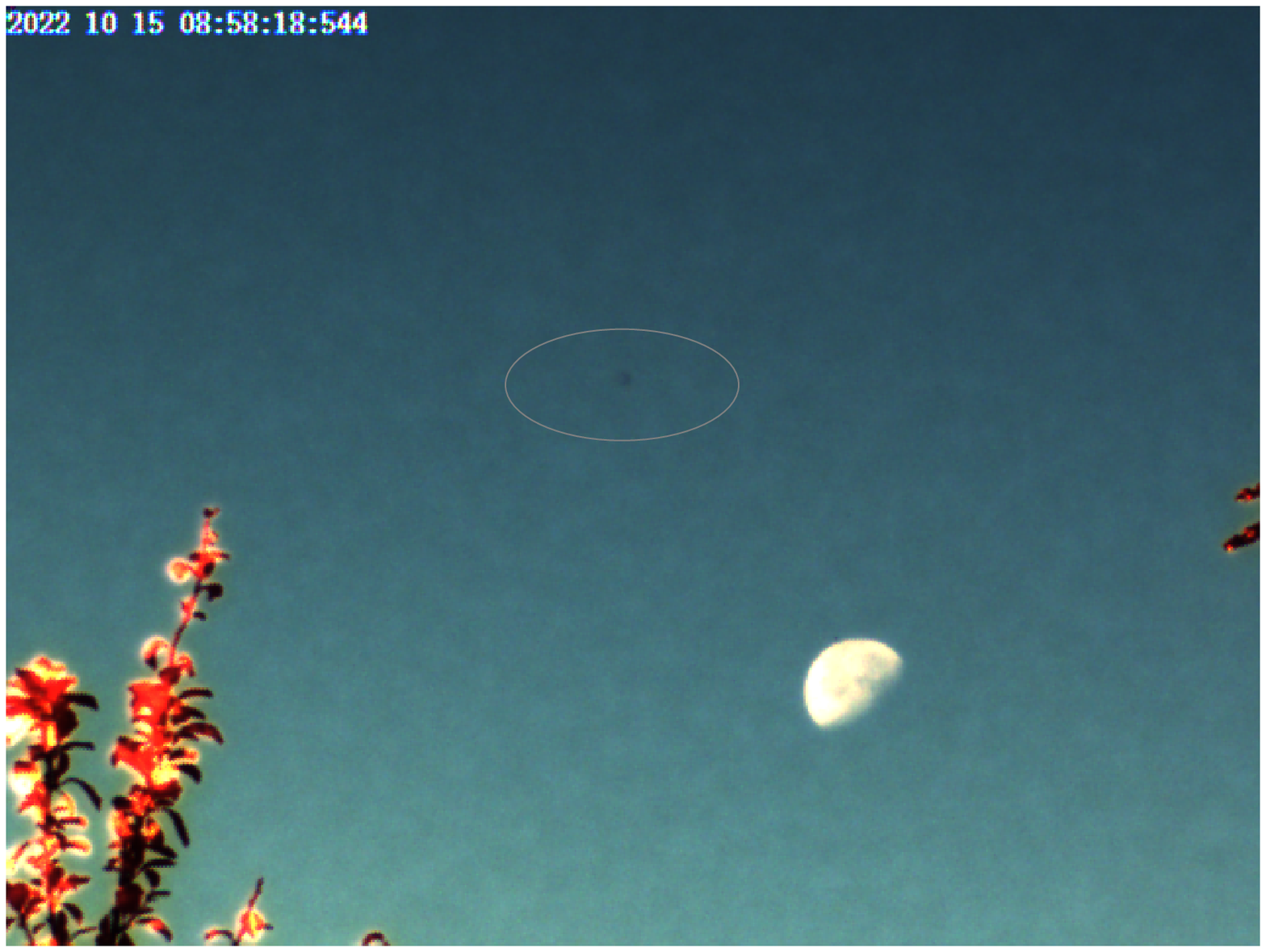,width = 0.9\linewidth} \caption{Arrival No. 5.}\label{fig1}
\end{minipage}
\hfill
\begin{minipage}[t]{.45\linewidth} 
\centering
\epsfig{file = 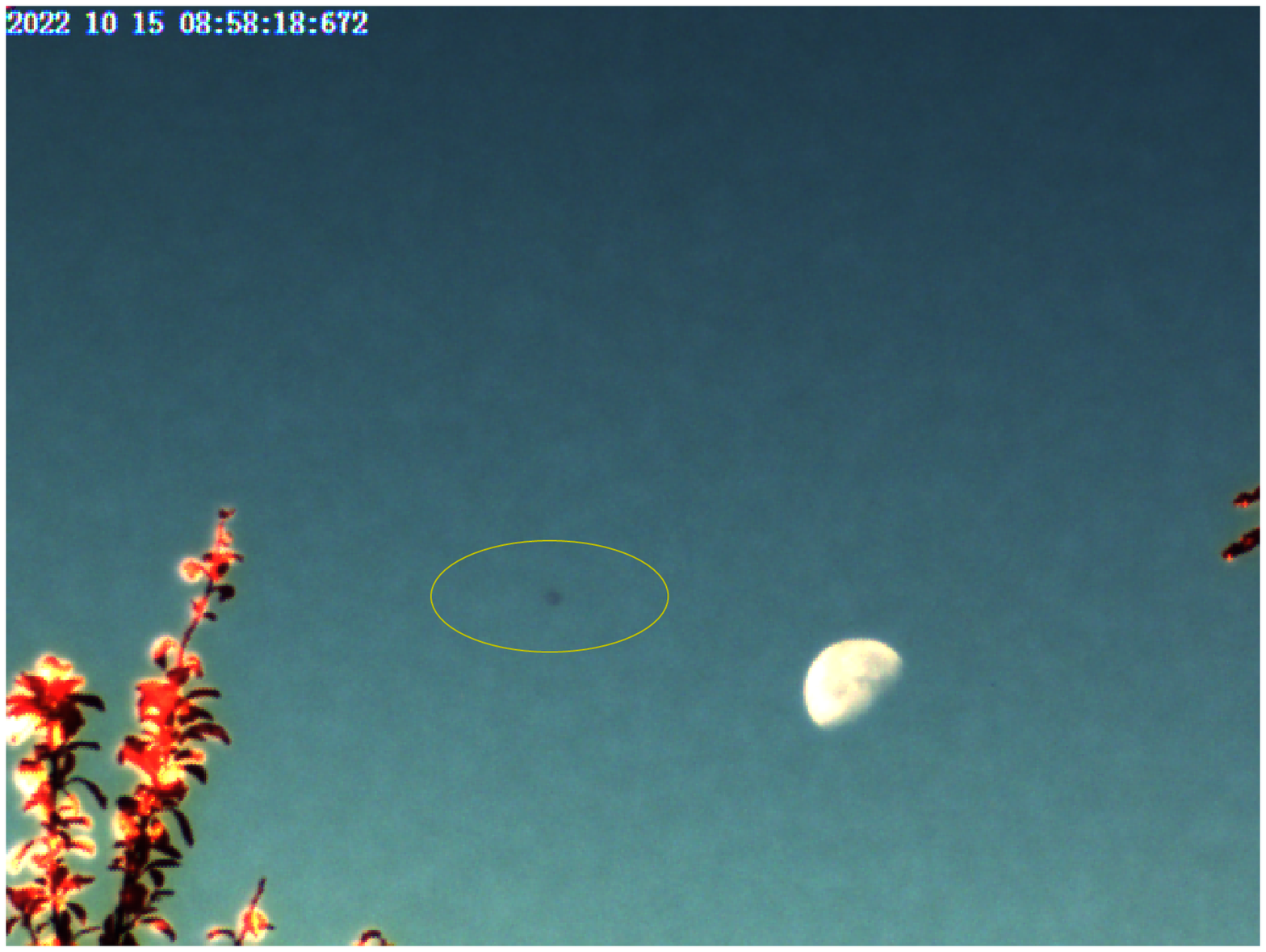,width = 0.9\linewidth} \caption{Arrival No. 5.}\label{fig2}
\end{minipage}
\end{figure}

\begin{figure}[!h]
\centering
\begin{minipage}[t]{.45\linewidth}
\centering
\epsfig{file = 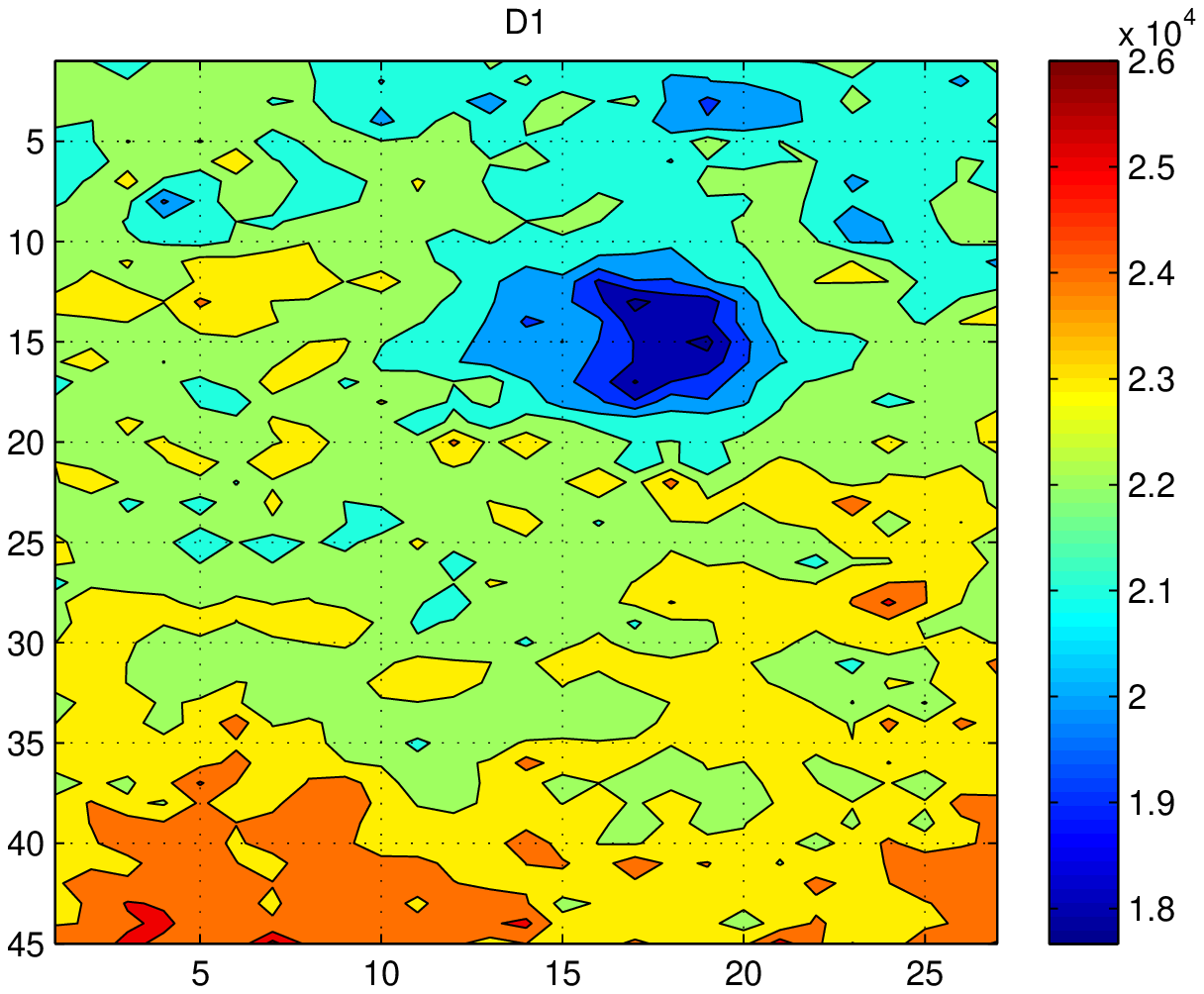,width = 0.9\linewidth} \caption{Arrival No. 5.}\label{fig1}
\end{minipage}
\hfill
\begin{minipage}[t]{.45\linewidth} 
\centering
\epsfig{file = 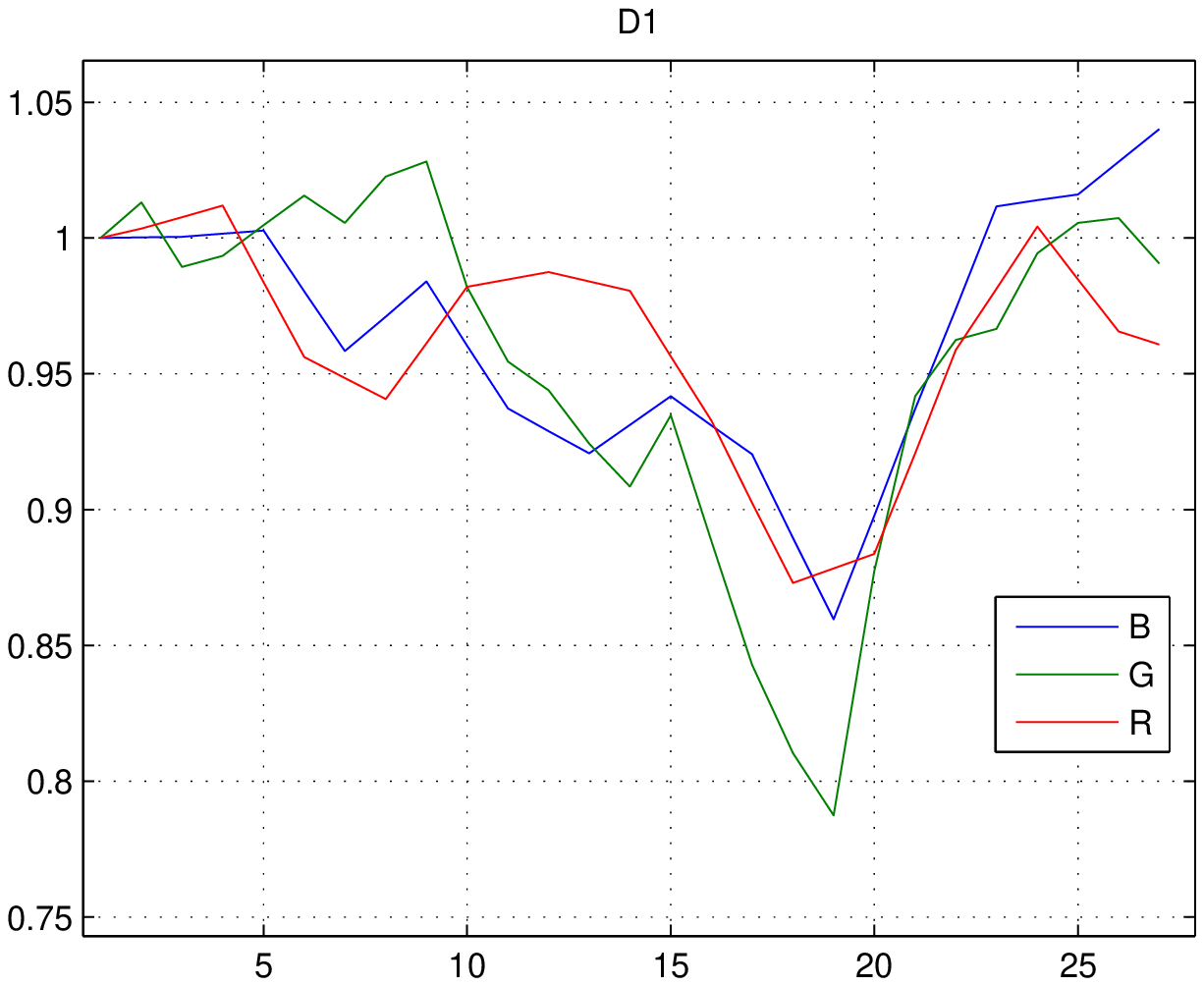,width = 0.9\linewidth} \caption{Color map.}\label{fig2}
\end{minipage}
\end{figure}


\section*{\sc Conclusions}

The Main Astronomical Observatory of NAS of Ukraine conducts a study of UAP. We used two meteor stations installed in Kyiv and in the Vinarivka village in the south of the Kyiv region. Observations were performed with color video cameras in the daytime sky. A special observation technique had developed for detecting and evaluating UAP characteristics.
There are two types of UAP, conventionally called Cosmics, and Phantoms. Cosmics are luminous objects, brighter than the background of the sky. Phantoms are dark objects, with contrast from several to about 50 per cent.

Two-site observations of UAPs at a base of 120 km with two synchronized cameras allowed the detection of two variable objects, at an altitude of 620 and 1130 km, moving at a speed of 256 and 78 km/s. 
Light curves of objects show a variability of about 10 Hz. Colorimetric analysis showed that the objects are dark: B - V = 1.35, V - R = 0.23.

We demonstrate the properties of several phantoms that were observed in Kyiv and the Kyiv region in 2018-2022. 
Phantoms are observed in the troposphere at distances up to 10 - 14 km. We estimate their size from 20 to about of 100 meters and speeds up to 30 km/s.

Color properties of bright flying objects indicate that objects are perceived as very dark. Albedo less than 0.01 would seem to make them practically black bodies, not reflecting electromagnetic radiation. We can assume that a bright flying object, once in the troposphere, will be visible as a phantom.

All we can say about phantoms is to repeat the famous quote: "Coming from the part of space, that lies outside Earth and its atmosphere. Means belonging or relating to the Universe".

\section*{\sc APPENDIX}

\subsection*{\sc Determination of distance to an object by colorimetry methods}

The colors of the object and the background of the sky make it possible to determine the distance using colorimetric methods. The necessary conditions are (1) Rayleigh scattering as the main source of atmospheric radiation; (2) and the estimated value of the object's albedo.
The scattered radiation intensity has the form:
\begin{equation}\label{}
I= I_{0} \cdot e^{-\sigma \cdot s}
\end{equation}
Here $s$ is the current distance, $\sigma $ is the Rayleigh scattering coefficient, and $I_{0} $ is the value of the intensity observed at sea level. The linear Rayleigh scattering coefficient $ \sigma$ has the form \cite{Allen}:
\begin{equation}\label{}
\sigma=3\cdot 10^{18} \cdot \delta \cdot (n-1)^{2} /\lambda^{4}/N
\end{equation}
Here $n$ is the refractive index of air, $ \lambda$ is the wavelength of light in microns, $ \delta$ is the depolarization coefficient equal to 0.97 for the Earth's atmosphere, and $N$ is the number of molecules in 1 $cm^{3}$ (Loshmidt number). Expression (7) can be written for both sky background and object. Expression (7) can be represented in stellar magnitudes as:
\begin{equation}\label{}
\Delta m = 1.086 \cdot \sigma \cdot (S-s_{obj})
\end{equation}
In the approximation of a homogeneous atmosphere with a height of 10 km, $S$ = 10/sin($h$), $h$ is the height of the object above the horizon, $s_{obj}$ is the distance to the object from the observer. This ratio is valid for a completely black body, with albedo equal to zero for any wavelength $ \lambda$. Thus, one can find the object's distance by measuring the difference between the stellar magnitudes of a phantom and the sky background.


\end{document}